\documentclass{emulateapj}

\slugcomment{Accepted to the {\it The Astrophysical Journal}}

\usepackage{color}
\usepackage{mathtools}
\usepackage{subfigure}
\usepackage{rotating}
\usepackage{epsfig} 
\usepackage{natbib}

\shorttitle{Accretion and Spots in DQ Tau}
\shortauthors{Bary \& Petersen}

\begin{document}


\title{Anomalous Accretion Activity and the Spotted Nature of the DQ Tau Binary System}


\author{Jeffrey S. Bary\altaffilmark{1,2} \& Michael S. Petersen\altaffilmark{1,3}}
\altaffiltext{1}{Colgate University, Department of Physics \& Astronomy,  13 Oak Drive, Hamilton, NY  13346}
\altaffiltext{2}{Visiting Astronomer, Laboratoire AIM Paris-Saclay, CEA/Irfu Universit\'{e} Paris-Diderot CNRS/INSU, 91191 Gif-sur-Yvette, France}
\altaffiltext{3}{University of Massachusetts, Department of Astronomy, 710 North Pleasant Street, Amherst, MA  01003}



\begin{abstract}

We report the detection of an anomalous accretion flare in the tight eccentric pre-main-sequence binary system DQ Tau. In a multi-epoch survey consisting of randomly acquired low to moderate resolution near-infrared spectra obtained over a period of almost ten years, we detect a significant and simultaneous brightening of four standard accretion indicators (Ca {\scshape ii} infrared triplet, the Paschen and Brackett series H {\scshape i} lines, and He {\scshape i} 1.083 $\mu$m), on back-to-back nights ($\phi$ = 0.372 \& 0.433) with the flare increasing in strength as the system approached apastron ($\phi$ = 0.5). The mass accretion rate measured for the anomalous flare is nearly an order of magnitude stronger than the average quiescent rate. While previous observations  established that frequent, periodic accretion flares phased with periastron passages occur in this system, these data provide evidence that orbitally-modulated accretion flares occur near {\it apastron}, when the stars make their closest approach to the circumbinary disk. The timing of the flare suggests that this outburst is due to interactions of the stellar cores (or the highly truncated circum{\it stellar} disks) with material in non-axisymmetric structures located at the inner edge of the circum{\it binary} disk.  We also explore the optical/infrared spectral type mismatch previously observed for T Tauri stars and successfully model the shape of the spectra from 0.8 to 1.0 $\mu$m and the strengths of the TiO and FeH bands as manifestations of large cool spots on the surfaces of the stellar companions in DQ Tau.  These findings illustrate that a complete model of near-infrared spectra of many T Tauri stars must include parameters for spot filling factors and temperatures.

\end{abstract}


\section{Introduction}

Since the initial discovery of a ``hot jupiter'' orbiting 51 Peg, exoplanet searches have firmly established that planetary systems, though unexpectedly diverse, are quite common in the Galaxy.   While the interaction between a forming planet and its circumstellar disk successfully explains orbital migration and the existence of ``hot jupiters," the similarly complex orbital dynamics in binary and higher-order multiple star systems might lead to the conclusion that planets should be rare in such systems.  Hence, initial searches for exoplanets selectively surveyed stars thought to be single star systems.  However, it has become apparent that planets do not form solely around single stars, as many exoplanet host stars have been revealed to be components of widely separated binary and higher-order multiple systems \citep[e.g.,][]{pati2002,egge2004,mugr2007,take2008,mugr2009}.  While these detections demonstrate that planets may form in the circumstellar disks orbiting individual stars in young multiple systems, Kepler 16b, 34b, 35b, and 47b and c indicate that planets also form in circumbinary disks orbiting tight binaries \citep{doyl2011,wels2012,oros2012}.

Observational surveys of main and pre-main-sequence solar-type stars have firmly established that more than half form in binary and multiple systems \citep[e.g.,][]{duqu1991,ghez1993,reip1993,lein1993,simo1995,duch2007}.  In fact, a notably higher multiplicity rate has been determined for pre-main-sequence stars suggesting that a substantial fraction of single field stars formed in multiple systems.  These results along with those of the exoplanet surveys, underscore the importance of studying the effects of multiplicity on the formation of stars and planetary systems.  Therefore, robust models of planet formation including the more dynamically complex multi-star systems with their combinations of circumstellar and circumbinary (circumsystem) material must be regarded as equally important for understanding planet formation and explaining the diversity of planetary systems.

The recent \emph{Kepler} detections of circum{\it binary} planets highlight the importance that understanding the dynamics of these complex systems will have on the study of star and planet formation.  The interactions of the stellar components with the circumbinary disks and the formation of planets within the circumbinary material must be considered when developing models of the complex star-disk interactions between the forming stars, their individual circumstellar disks, and the circumbinary(system) disk.  

At present, hydrodynamic simulations of the complex interactions between binary protostars with circumstellar and circumbinary disks lay a foundation for characterizing the impact of multiplicity on the formation and evolution of planets in such systems \citep{arty1991,arty1994,arty1996,rozy1997,gunt2002,pier2008a,pier2008b,hana2010,deva2011}.  These simulations depict how the central sources in a binary system each truncate the outer region of the companion's circumstellar disk, while quickly clearing a dynamically unstable region located between the central stars and the inner region of the circumbinary disk.  Once the inner disks are cut off from the larger reservoir of circumbinary material, the truncated inner disks will accrete onto the central star on a timescale expected to be too short to allow for the formation of a planetary system.  Observations of scattered light and continuum emission from dust grains in the circumstellar and circumbinary disks have confirmed the existence of nearly dust-free gaps in several nearby binary systems \citep[i.e., GG~Tau, SR~24, DQ~Tau;][]{dutr1994,rodd1996,jens1997,mcca2002,kris2002,kris2005,piet2011}.   

Simulations of young binary systems also provide a physical understanding of how circumbinary disk material falls into the unstable region via ``high-velocity streams" of gas and dust \citep{arty1996,rozy1997,gunt2002,deva2011}.  The streams of material, also referred to as ``streamers," will fall towards the stellar companions either replenishing the inner circumstellar disks or perhaps accreting directly onto the surfaces of the stellar components.  In the cases of SR~24 and GG~Tau, scattered light ({\it H-}band) and millimeter continuum observations provide evidence that the unstable regions are not completely devoid of material \citep{maya2010,piet2011}; these observations lend support to the theory that predicts accretion streams.  In fact, both hint at the presence of narrow streams of gas and dust flowing from the circumbinary disk to the inner circumstellar disks.  For the GG~Tau system, known to possess a distinctive circumbinary ring with an inner radius of $\sim$180~AU, a recent high spatial resolution image of molecular hydrogen emission shows a ridge of shock-excited gas located within the circumbinary disk and near the edge of the truncation region of the inner circumstellar disks \citep{beck2012}.  The H$_2$ emission is spatially coincident with the location of the dust streamers and appears to confirm the existence of streamers in the GG~Tau~A system.

\subsection{The DQ Tau System}

\begin{figure*}[ht]
\begin{center}
\includegraphics[angle=0,width=1.8\columnwidth]{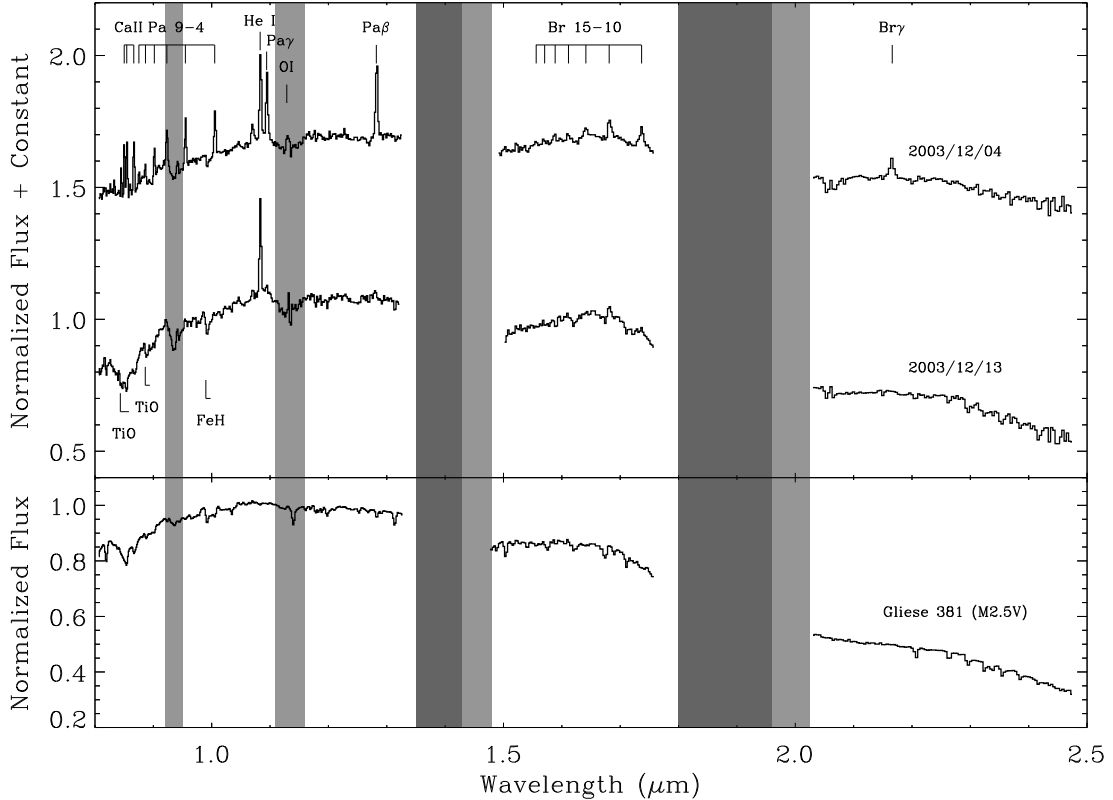}
\caption{Two epochs of normalized NIR spectra of DQ~Tau acquired with CorMASS at the Vatican Advanced Technology Telescope.  The observation taken on 2003 December 4 UT captured the system during an outburst phase as indicated by the presence of strong H~{\scshape i} and Ca~{\scshape ii} infrared triplet emission lines.  The lower spectrum 2003 December 13 UT, obtained eight days after the spectrum above, shows little evidence of H~{\scshape i} or Ca~{\scshape ii} emission suggesting that relatively little accretion activity is occurring during this point in the orbital phase of the system.  The bottom panel shows the spectrum of an M2.5V spectral standard (Gliese 381) from the SpeX library \cite{rayn2009}.  The resolution of the standard star spectrum has been degraded to match the resolution of CorMASS.  Following the plot format from \citet{vacc2011}, regions of poor atmospheric transmission are identified by the gray ($<$80\%) and dark gray ($<$20\%) vertical boxes.}
\label{fig:sample}
\end{center}
\end{figure*}

The DQ~Tau system, the subject of the study presented here, is a tight binary \citep[{\it P}~$\approx$~15.8 days;][]{huer2005}  located in the nearby Taurus-Auriga star forming region (D~$\sim$~140~pc; Age~$\sim$~1-3 Myr).  DQ~Tau possesses two nearly equal mass stars ($M_a+M_b$~$\sim$~1.3~M$_\odot$) with a high orbital eccentricity \citep[{\it e}~=~0.556;][]{math1997}.  For this binary system, possessing a semi-major axis of $\sim$~0.13~AU, the inner truncation radius of the circumbinary disk should reside at $\sim$~0.4~AU.  The spectral energy distribution (SED) shows both near- and mid-infrared excesses suggesting the presence of both a circumbinary disk and warm dust in the environment of the circusmtellar disks \citep{stro1989,skru1990}.  Recent Keck Interferometer observations and modeling of the {\it K-}band excess emission confirm that the warm dust must exist in the region where circumstellar disk material should reside with an extent of 0.1-0.2~AU \citep{bode2009}.  Combined with previous detections of warm CO line emission \citep[T$\sim$1200~K; R$\sim$0.1~AU;][]{carr2001}, these observations suggest that the dynamically cleared region located within the circumbinary disk contains dust and gas, leading to the conclusion that material is continually drawn inward from the outer disk.

%These observations follow a detection of CO fundamental rovibrational emission requiring an excitation temperature of $\sim$1200~K, locating the gas to within 0.1~AU and very likely within the cleared gap in the DQ~Tau system.  These observations confirm the hydrodynamic model predictions that material from the circumbinary disk can flow into the unstable region, likely replenishing the inner circumstellar disks.

For high-eccentricity binary systems such as DQ~Tau, hydrodynamic models predict that orbitally-modulated accretion activity should result in accretion flares occurring at periastron passes of the stellar companions \citep{gunt2002,deva2011}.  Quasi-periodic flares phased with periastron in the DQ~Tau system were first observed, using broad-band photometry, by \citet{math1997}.  Subsequent multi-epoch optical and near-infrared spectroscopic observations detected increases in the fluxes of several standard accretion signatures similarly phased with periastron passages in DQ~Tau and two other tight binaries, UZ~Tau~E and AK~Sco \citep{basr1997,huer2005,jens2007,bary2008}.  For these systems, in which the stellar companions come within a few stellar radii of one another, the flaring could be explained by interacting magnetospheres and are likely the cause of millimeter and X-ray flares  observed in the DQ~Tau and UZ~Tau~E systems \citep{salt2008,salt2010,getm2011,kosp2011}.  As the stars recede, the strength of the interaction between the magnetospheres diminishes rapidly with increasing distance.  The length of the millimeter and X-ray flares is on the order of hours, while the optical photometric and spectral feature flares last for several days or as much as a third of the orbit.  Noting that the longer flares occur relatively far in orbital phase from periastron passage, \citet{jens1997} argue that accretion activity is most likely responsible for this flaring.  Therefore, these observations likely confirm the hydrodynamical model predictions that material from the circumbinary disk can flow into the unstable region, replenishing the inner circumstellar disks.

For widely-separated young binary systems, accretion streams provide a mechanism for increasing the lifetimes of the individual circumstellar disks, potentially impacting the formation of planets in such systems.  While it may be unreasonable to expect planets to form on stable orbits in the tidally-truncated circumstellar disks associated with the stellar companions in the DQ~Tau system, the recent \emph {Kepler} detections of circumbinary planets clearly indicate that the circumbinary disk is a viable site for planet formation.  While star-disk interactions play an important role in the evolution of planet-forming disks in single star systems, star-star-circumbinary disk interactions are presumed to be equally important in the evolution of circumbinary planetary systems.  In addition, as described by \citet{arty1996} and \citet{rozy1997}, the structure of a tight binary that is in the process of clearing a gap in the inner region of its circum{\it binary} disk is analogous to the structure of a transitional disk system in which a gas giant planet is suspected of clearing a gap in the inner region of a circum{\it stellar} disk.  The potential for analogous orbitally-modulated accretion activity in transitional disk systems underscores the importance of studying young binary systems and their star-star-disk interactions.

In this paper, we present a multi-epoch spectroscopic study of pulsed-accretion activity in the DQ~Tau system.  We use near-infrared (IR) spectral accretion signatures (i.e., H~{\scshape i}, Ca~{\scshape ii}, and He~{\scshape i}) to establish an average mass accretion rate for the system in its quiescent accretion phase.  We report for the first time the detection of an anomalous accretion flare well outside of the usual periastron related activity, timed more closely in orbital phase with an apastron passage of the system.  The detection is discussed in the context of current hydrodynamic models of star-star-disk-disk interactions.  The moderate resolution spectra allow for the determination of the infrared spectral type of the DQ~Tau system based on the spectral shape, molecular absorption features, and metallic absorption lines.  Similar to previous studies of near-IR spectra of T Tauri stars (TTSs), we find a discrepancy between the spectral types previously derived from optical spectroscopy and the near-IR spectral type.  Other authors have also noted a color anomaly at IR wavelengths for TTSs and have suggested that both discrepancies may be related to the existence of large cool spots on the surfaces of these stars \citep{gull1998,gull1998b}.  We construct simple synthetic spectra to represent a spotted star from spectral standards and perform a fit to a representative DQ~Tau spectra.  We solve simultaneously for visual extinction, photospheric temperature, spot temperature, and spot filling factor by fitting the strong TiO and FeH bands and the shape of the near-IR spectrum.  The best fits cover a small region of parameter space and demonstrate that the existence of large, cool spots are reasonable explanations for the observed optical/IR spectral mismatch and the color anomaly.  
 
\section{Observations and Data Reduction}

\begin{figure*}[ht]
\begin{center}
\includegraphics[angle=0,width=1.8\columnwidth]{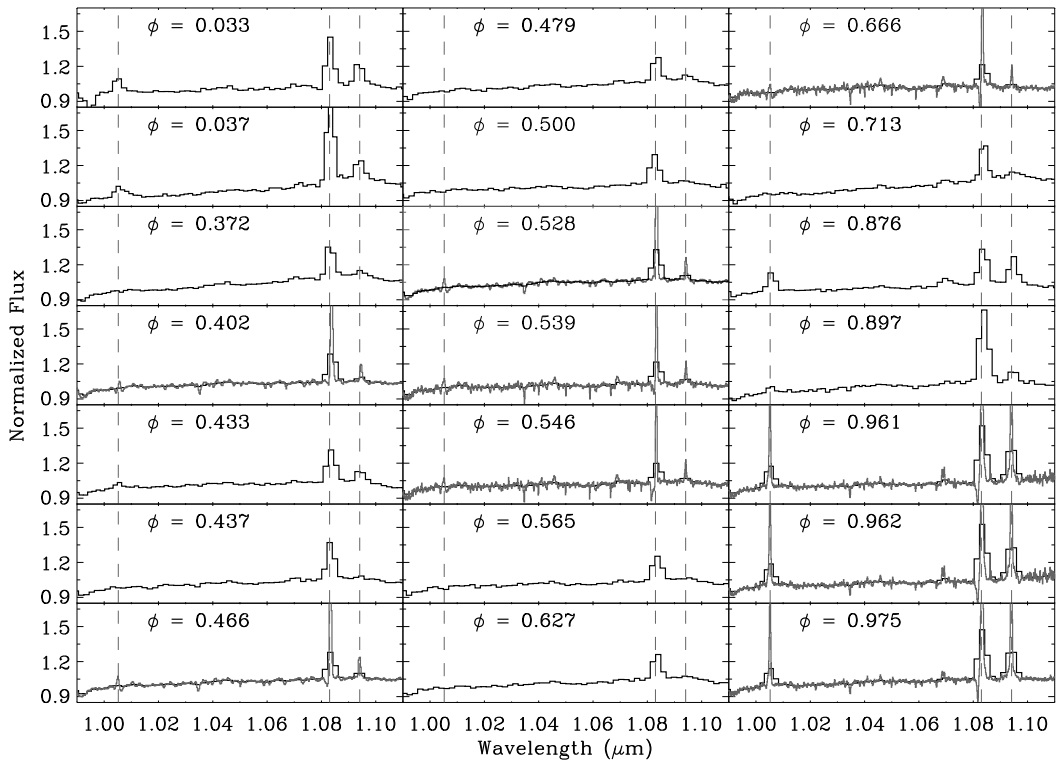}
\caption{Plotted are 18 epochs of DQ~Tau spectra from 1.07 to 1.15 $\mu$m to highlight the variations of the He~{\scshape i} and Pa$\gamma$ features at 1.083~$\mu$m and 1.0938~$\mu$m, respectively.  For the epochs where SpeX and TripleSpec data of higher resolution were collected, we present both the full resolution spectrum (bold) and a spectrum degraded to the resolution of the CorMASS data (gray).  The higher resolution spectra containing the He~{\scshape i} feature show how the blueshifted absorption feature waxes and wanes and changes shape.  In addition, the degraded spectra demonstrate the utility of the high resolution spectra for measuring the quiescent accretion activity and the unusual strength of the Pa$\gamma$ feature associated with the potential apastron flare at $\phi$ = 0.372 and 0.433.}
\label{fig:mpag}
\end{center}
\end{figure*}

\begin{figure*}[ht]
\begin{center}
\includegraphics[angle=0,width=1.8\columnwidth]{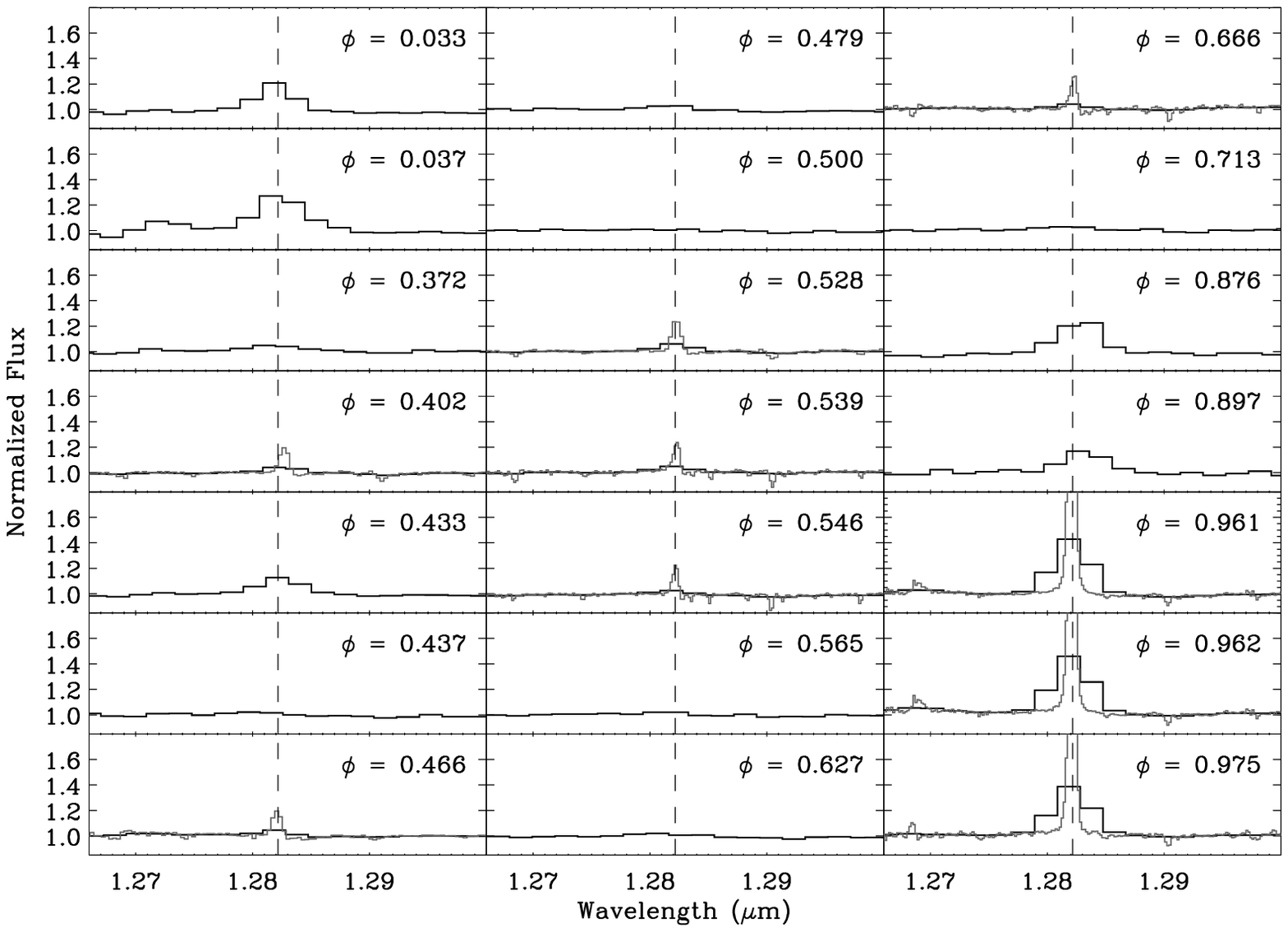}
\caption{Plotted are 18 epochs of DQ~Tau spectra from 1.266 to 1.30 $\mu$m to highlight the variations in the Pa$\beta$ feature at 1.2818 $\mu$m.  For the epochs where SpeX and TripleSpec data of higher resolution were collected, we present both the full resolution spectrum (bold) and a spectrum degraded to the resolution of the CorMASS data (gray).  The higher resolution spectra illustrate the changes in the strength and shape of the Pa$\beta$ feature, including the appearance of a redshifted absorption component to the line in the $\phi$~=~0.666 observation.  The degraded spectra demonstrate the utility of the high resolution spectra for measuring the quiescent accretion activity and the unusual strength of the Pa$\beta$ feature associated with the potential apastron flare at $\phi$ = 0.372 and 0.433.}
\label{fig:mpab}
\end{center}
\end{figure*}

The multi-epoch data presented in this study were collected over the past ten years using three different instruments (CorMASS, TripleSpec, and SpeX) mounted on three different telescopes.  The initial observations were made as part of a larger spectroscopic variability study of near-IR emission lines observed from actively accreting TTSs \citep{bary2008}.  The survey utilized CorMASS \citep[Cornell-Massachusetts Slit Spectrograph;][]{wils2001}, a low-resolution (R$\sim$300) cross-dispersed near-IR spectrograph that provides continuous wavelength coverage from 0.8 to 2.5~$\mu$m.  Data were collected on the 1.8-m Vatican Advanced Technology Telescope (VATT) atop Mt.\ Graham in Arizona from 2003 December through 2005 January.  In 2005, CorMASS was moved to Apache Point Observatory (APO) in New Mexico where it was positioned at the Naysmith focus of the 3.5-m Astrophysical Research Consortium (ARC) telescope.  The survey continued as spectra were collected at APO in late 2005 and early 2006.  As part of the larger variability study, CorMASS observations of DQ~Tau were not phased with the orbital period.  Instead, they were collected randomly when telescope time was available and the targets were well positioned.  Additional observations of the DQ~Tau system were collected during 2010 January and 2012 October observing runs with TripleSpec \citep[TSpec;][]{wils2004}, also mounted on the ARC telescope at APO.  An instrumental cousin of CorMASS, TSpec is a cross-dispersed near-IR spectrograph providing continuous wavelength coverage from 0.9 to 2.5~$\mu$m with an order of magnitude higher resolving power (R$\sim$3,000).  In addition, three consecutive nights of data were acquired at the IRTF during 2011 November, using SpeX, also a cross-dispersed near-IR spectrograph providing nearly continuous wavelength coverage from 0.8 to 2.4~$\mu$m in the short-wavelength cross-dispersed mode (SXD; R$\sim$2000).  Table~1 summarizes the observations including the orbital phases at which the measurements were taken.

The slit dimensions are 1$\farcs$6 and 0$\farcs$75 in width and 12$\farcs$2 and 5$\farcs$7 in length for CorMASS on the VATT and APO, respectively.  For TSpec at APO and SpeX at IRTF, the dimensions are 1$\farcs$1 and 0$\farcs$5 in width and 45$\arcsec$ and 15$\arcsec$ in length, respectively.  Observations of both DQ~Tau and telluric calibration sources with CorMASS+VATT, TSpec+APO, and SpeX+IRTF were made using a standard ABBA nod pattern, allowing for an efficient use of telescope time and accurate removal of thermal background, sky emission lines, and dark current.  The small slit length for CorMASS+APO required target data to be accompanied by a sky observation of equal exposure time, in which the telescope was nodded to a blank sky position within a few arcseconds of the target.  While the object + sky observations are slightly less efficient than nodding, both techniques allow for accurate removal of background, sky lines, and dark current.

Data reduction was performed using the IDL Spextool package and its adaptations for CorMASS and TSpec, all designed to process cross-dispersed data \citep{cush2004}.  Each spectrum was wavelength calibrated using resolved OH telluric emission features.  HD~26710, a G2V star was used as the telluric standard for most CorMASS observations.  In two epochs, an observation of an alternate telluric calibrator gave better results (see Table~\ref{tbl:1} for details).  For CorMASS observations, H~{\scshape i} Paschen and Brackett absorption features present in the telluric spectrum were removed by linear interpolation between the continuum points that bracket the feature.  Corrected spectra were then multiplied by a blackbody function matching the temperature of the telluric calibration star to preserve the continuum shape of the DQ~Tau spectra.  For TSpec and Spex observations, final telluric correction was performed using an A0V star, and the telluric correction procedure built-in to the SpeX and TSpec reduction package.  This procedure creates a telluric correction spectrum by generating a synthetic spectrum of the A0V calibrator star based on a model spectrum of Vega.  The resulting synthetic A0V spectrum of the calibration star is then divided by the observed spectrum to remove H~{\scshape i} features and weak metal lines.  The target star spectrum is then multiplied by this ``telluric correction spectrum'' removing telluric features and the instrumental response function, while preserving the photospheric features and shape of the spectral energy distribution \citep[for details see][]{vacc2003}.  

Figure~\ref{fig:sample} presents two epochs of telluric corrected DQ~Tau spectra over the full wavelength coverage obtained with CorMASS on the VATT.  The two spectra were obtained eight days apart on 2003 December 4 and 13 UT  with orbital phases, $\phi$ equal to 0.876 and 0.437, respectively.  In the spectrum taken on December 4, strong emission features associated with the Paschen and Brackett H~{\scshape i} series and the Ca~{\scshape ii} infrared triplet are clearly detected.  For comparison, the spectrum of a M2.5V star obtained from the IRTF Spectral Library \citep{rayn2009} is included with its resolution degraded to match that of CorMASS.

\begin{deluxetable}{lccc}
\tabletypesize{\scriptsize}
\tablecaption{DQ~Tau Observations \label{tbl:1}}
\tablewidth{0pt}
\tablehead{
				\colhead{UT Date}
			&  	\colhead{Telescope+Instr.}
			&  	\colhead{Orbital}
			&  	\colhead{Telluric}  \\
				\colhead{} 
			&	\colhead{}
			&	\colhead{Phase\tablenotemark{a}}
			&	\colhead{Calibrator}}
\startdata
2003/12/04   &   VATT+CorMASS  & 0.876 $\pm$ $^{0.027}_{0.082}$& HD 26710  \\
2003/12/13  &   `` "  		   	&   0.437 $\pm$ $^{0.028}_{0.084}$  & `` " \\
2003/12/14   &   `` "  		   	&0.500 $\pm$ $^{0.028}_{0.084}$  & `` " \\
2003/12/15   &   `` " 		   	& 0.565 $\pm$ $^{0.028}_{0.084}$ & `` " \\
2003/12/16  &  `` "  		   	& 0.627 $\pm$ $^{0.028}_{0.084}$   & `` " \\
2005/01/29  &   `` " 		  	&  0.479 $\pm$  $^{0.031}_{0.093}$   & `` " \\
2005/10/16  &   APO+CorMASS  &  0.033 $\pm$ $^{0.031}_{0.093}$ & HIP 20899\\
2006/01/15  &   `` " 		  	& 0.713 $\pm$ $^{0.034}_{0.102}$  & HD 26710 \\
2006/01/18  &   `` " 		  	& 0.897 $\pm$ $^{0.034}_{0.102}$      & `` " \\
2006/01/19 &   `` "  		  	& 0.037 $\pm$  $^{0.034}_{0.102}$  & `` " \\
2006/10/04 &   `` "  		  	& 0.372 $\pm$ $^{0.035}_{0.107}$ & `` " \\
2006/10/05 &   `` "  		  	& 0.433 $\pm$  $^{0.034}_{0.102}$   & HIP 19767 \\
2010/01/04 &   APO+TSpec  	  & 0.539 $\pm$  $^{0.045}_{0.135}$ & HD 283558 \\
2010/01/04 &   `` "  		  	& 0.546 $\pm$  $^{0.045}_{0.135}$   & `` " \\
2010/01/06 &   `` "  		 	& 0.666 $\pm$ $^{0.045}_{0.135}$  & `` "  \\
2011/11/12 &   IRTF+SpeX		 & 0.402 $\pm$ $^{0.048}_{0.139}$  & HD24555  \\
2011/11/13 &   `` "		 	& 0.466 $\pm$ $^{0.048}_{0.139}$   & HD34203 \\
2011/11/14 &   `` "		  	& 0.528 $\pm$ $^{0.048}_{0.139}$  & `` "  \\
2012/10/02 &  APO+TSpec 	&0.961 $\pm$ $^{0.051}_{0.148}$  			&HD37887\\
2012/10/02 & `` "			&	0.962 $\pm$ $^{0.051}_{0.148}$ 		&`` '' \\
2012/10/02 & `` "			&	0.975 $\pm$ $^{0.051}_{0.148}$ 		&`` " \\
\enddata

\tablenotetext{a}{Orbital phases were calculated beginning on Julian date 2449582.54 using the orbital period of 15.8016$\pm$$^{0.002}_{0.006}$ measured by \citet{huer2005}.}

\end{deluxetable}

\section{Results}

\begin{figure*}[ht]
\begin{center}
\includegraphics[angle=0,width=1.8\columnwidth]{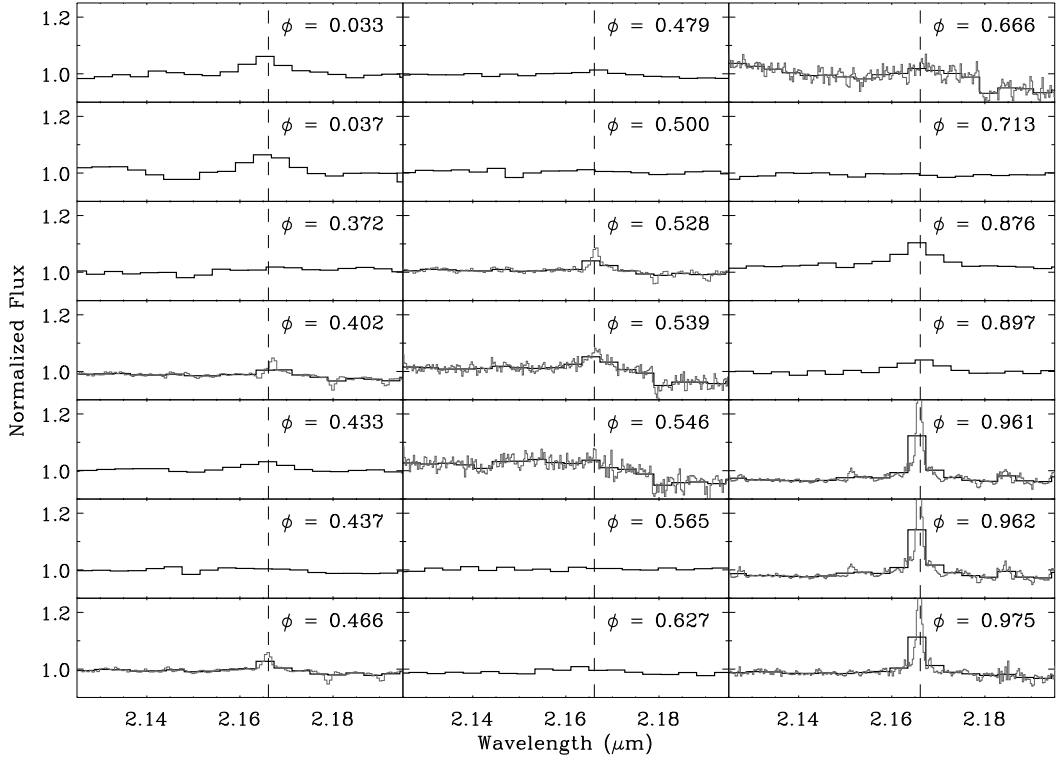}
\caption{Plotted are 18 epochs of DQ~Tau spectra from 2.125 to 2.195~$\mu$m to highlight the variations in the Br$\gamma$ feature at 2.1655 $\mu$m.  For the epochs where SpeX and TripleSpec data of higher resolution were collected, we present both the full resolution spectrum (bold) and a spectrum degraded to the resolution of the CorMASS data (gray).  The higher resolution spectra illustrate the changes in the strength of the Br$\gamma$ feature.  In the few moderate resolution spectra, we note the lack of a redshifted absorption feature.  The degraded spectra demonstrate the utility of the high resolution spectra for measuring the quiescent accretion activity and the unusual strength of the Br$\gamma$ feature associated with the potential apastron flare at $\phi$ = 0.372 and 0.433.}
\label{fig:mbrg}
\end{center}
\end{figure*}

The multi-epoch low- and moderate-resolution near-IR spectra collected over the past ten years for this project provide the opportunity to study several different aspects of the DQ~Tau system.  We first present measurements of the mass accretion rate based on the strengths of the H~{\scshape i} emission lines.  Using the eleven observations that do not show any flaring signature, we determine an average quiescent mass accretion rate for the system and use this to establish the existence of an anomalously strong accretion flare located near one of the apastron passes.   {Next, as a probe of possible different modes of accretion activity in the system, we compare the H~{\scshape i} line ratios during outburst and quiescent phases and use those to independently determine the temperature and density of the accreting gas in both accretion regimes using a new line excitation model \citep{kwan2011}.  In the next section, we present an analysis of the spectral shapes and molecular and atomic photospheric absorption features resulting in a determination of the infrared spectral type for DQ~Tau.  The discrepant IR spectral type leads us to formulate a simple spot model that successfully replicates the TiO and FeH band features and offers a coherent explanation for both the optical/IR spectral type discrepancies and color anomalies observed for TTSs.  Finally, we place an upper limit on the emission from the $v~=~1-0$~S(1)~H$_2$ feature at 2.12~$\mu$m, which we discuss in the context of shocks associated with accretion ``streamers" falling inward from the circumbinary disk. } 

\subsection{Accretion Flare Indicators}

\begin{figure}[ht]
\begin{center}
\includegraphics[angle=90,width=1.0\columnwidth]{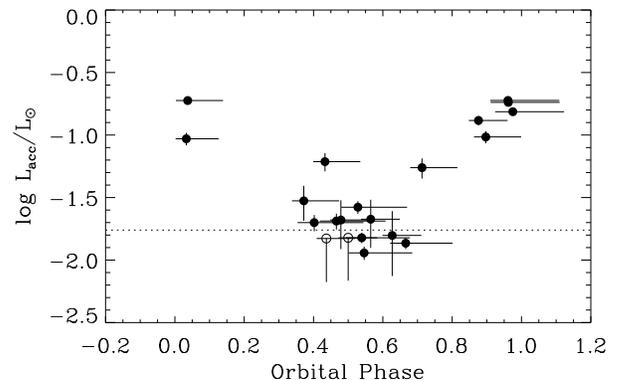}
\caption{Plotted are the values of the accretion luminosity determined from the Pa$\beta$ line luminosities using the Muzerolle relations as a function of orbital phase.  Filled circles represent detections, while open circles designate upper limits.  Error bars on the values of $L_{acc}$ are strictly measurement uncertainties.  Note that some error bars are smaller than the symbol size.  Error bars on the orbital phases are derived from \citet{huer2005} uncertainties in the orbital period and are propagated through time using periastron on JD 2449582.54 as the origin.  The dashed horizontal line indicates the average value of $L_{acc}$ measured during the quiescent portion of the orbit.}
\label{fig:lacc}
\end{center}
\end{figure}

Optical and near-IR hydrogen emission features (i.e., Balmer, Paschen, and Brackett series) are defining characteristics of the classical T Tauri classification of pre-main-sequence stars.  These features are accepted as signposts for the presence of circumstellar material and indicators of accretion activity.  In the paradigm of magnetospherically-guided accretion, matter free-falls from the circumstellar disk toward the stellar surface guided by the large-scale dipole-like stellar magnetic field.   The material slams into the stellar photosphere, releasing energy as a shock front forms a hot spot on the surface of the star, heating both the shocked gas and the trailing column of infalling material.  Ionization and recombination of hydrogen in the dense gas flows are thought to be responsible for the majority of flux produced in the H~{\scshape i} emission lines present in the spectra of TTSs.  For some T Tauri systems, recent observations have demonstrated that a small fraction of the hydrogen emission is spatially extended and associated with outflowing gas \citep{beck2010,coff2010,davi2011}.  While the contribution of emission from the outflows to the integrated line flux is negligible for most T Tauri systems, it is potentially more important for less evolved, embedded Class~I systems.  Given the relationship between accretion and outflows and the fact that many observational studies have shown that emitting H~{\scshape i} gas correlates well with other tracers of accreting gas, it is reasonable to conclude that H~{\scshape i} line emission remains a trusted proxy for determining mass accretion rates in TTS systems.  The wavelength coverage of CorMASS, TSpec, and SpeX and multiple epochs of observations provides the opportunity to study several hydrogen transitions simultaneously in the Paschen and Brackett series and to measure the variations in mass accretion rates as a function of orbital phase.

In addition to the H~{\scshape i} features, the He~{\scshape i}~1.083~$\mu$m feature is another well-studied spectral characteristic of many TTSs that falls within the wavelength coverage of CorMASS, TSpec, and SpeX \citep[e.g.,][]{beri2001,dupr2005,edwa2006,fisc2008}.  At moderate resolution, this feature may appear with a P Cygni and/or inverse P Cygni profile, possessing either a blue-shifted or a red-shifted absorption component, or both.  Models of this feature describe the formation in either accreting and/or outflowing gas with moderate success \citep{kwan2007,fisc2008}.  We monitor the strength of this feature and the appearance of absorption components as a potential indicator of differing modes of accretion activity in the DQ~Tau system.

Lastly, the Ca~{\scshape ii} infrared triplet near 0.85~$\mu$m has also been observed to correlate with accretion activity \citep{muze1998a,azev2006} and falls within the wavelength coverage of CorMASS and SpeX.  We will assume that the presence of the H~{\scshape i} and Ca~{\scshape ii} emission features directly indicate ongoing mass accretion in the DQ~Tau system.  

\subsection{Hydrogen Line Variability}

\begin{figure}[ht]
\begin{center}
	\includegraphics[angle=0,width=0.9\columnwidth]{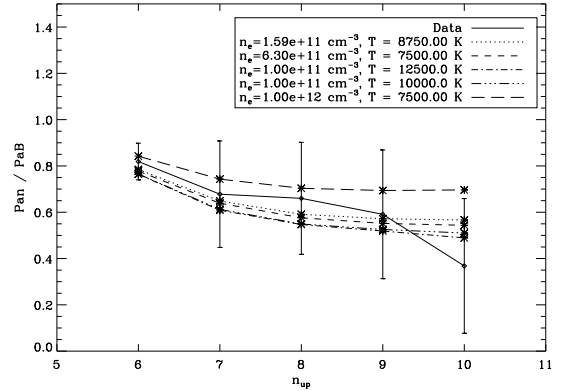}
	\caption{The Paschen decrement plotted for an observation of DQ~Tau on 2003 December 04 UT and includes five Paschen series lines with detectable line fluxes.  The observed line ratios are open diamonds connected by a solid line.  The five best KF model fits are overplotted and ordered in the figure legend by goodness-of-fit.  These models tightly constrain the density, yet do not distinguish between the temperatures in the KF models.}
	\label{fig:pantopaba}
\end{center}
\end{figure}

\begin{figure}[ht]
\begin{center}
	\includegraphics[angle=0,width=0.9\columnwidth]{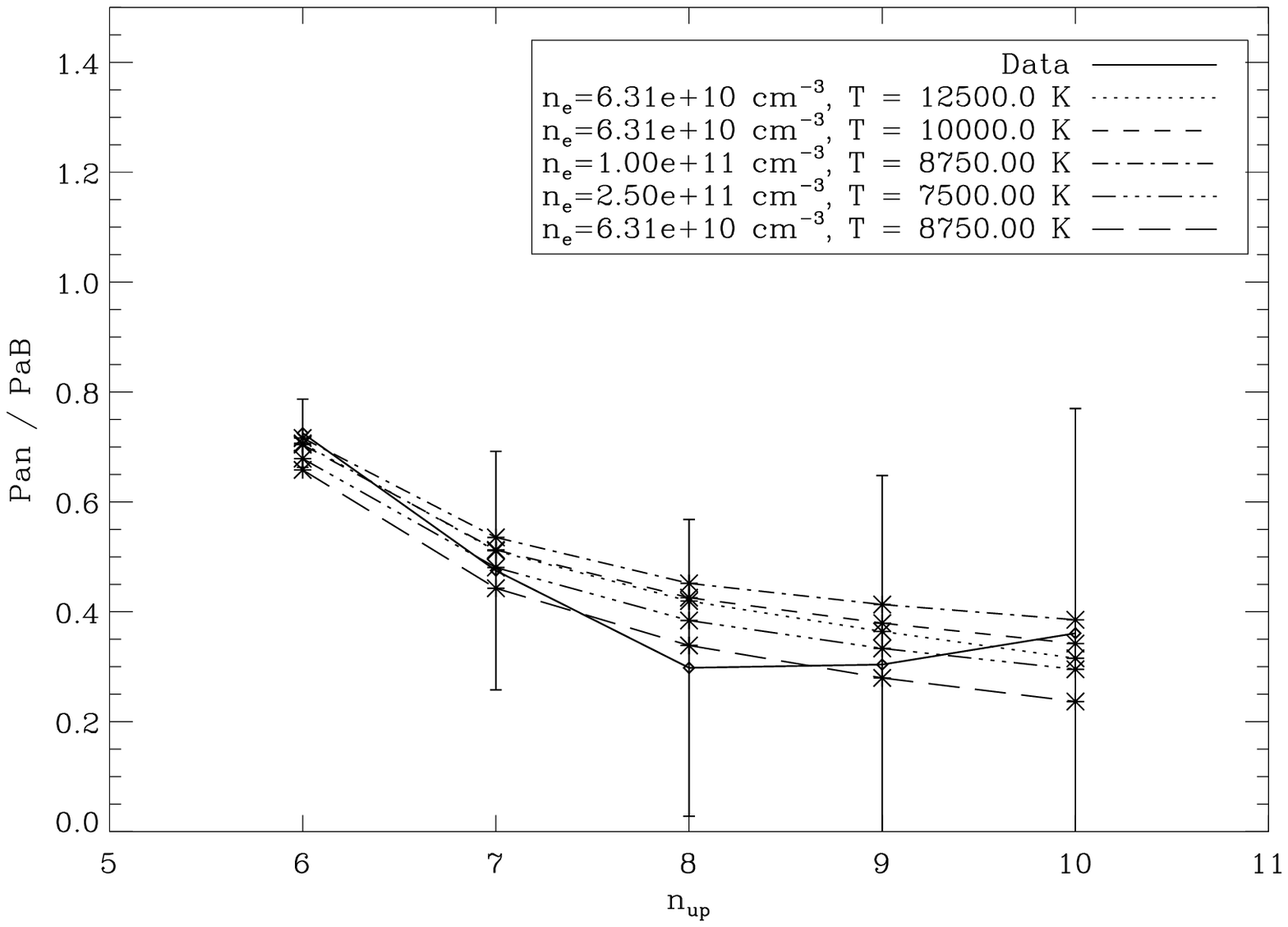}
	\caption{Same as Figure~\ref{fig:pantopaba} for observation of DQ~Tau on 2006 October 05 UT.}
	\label{fig:pantopabb}
\end{center}
\end{figure}

\begin{figure}[ht]
\begin{center}
	\includegraphics[angle=0,width=0.9\columnwidth]{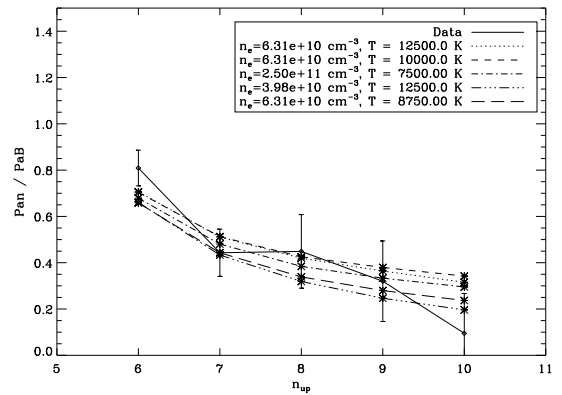}
	\caption{Same as Figure~\ref{fig:pantopaba} for observation of DQ~Tau on 2012 October 02 UT.}
	\label{fig:pantopabc}
\end{center}
\end{figure}

Figure~\ref{fig:mpag} shows the variations in strength of both the He~{\scshape i} and Pa$\gamma$ emission features.  Notably, the He~{\scshape i} feature remains strongly in emission throughout the entire orbit of the system with only moderate variations in strength.  The moderate resolution spectra available at $\phi$~=~0.402, 0.466, 0.528, 0.539, 0.546, and 0.666 show that the structure of the feature undergoes some interesting variations.  For instance, the appearance of the blue-shifted absorption feature at $\phi$~=~0.546 and 0.666 suggests the presence of a {\it variable} outflowing wind ($v$~$\sim$~200-300~km~s$^{-1}$) in the system since the feature is much weaker or completely absent at the other orbital phases.  The presence of the blue-shifted absorption component in the He~{\scshape i} feature observed in a few epochs of the moderate resolution spectra suggest that some of the variation observed in the low-resolution spectra are correlated with the presence or absence of the blue-shifted absorption component.  The Pa$\gamma$ feature becomes undetectable in two epochs of the low-resolution data.  As expected for orbitally-modulated accretion, the Pa$\gamma$ feature appears strongest in epochs near periastron.  The TSpec and SpeX spectra collected far enough from periastron  allow us to confidently detect this feature during the quiescent phase of the orbit.  

Figures~\ref{fig:mpab} and \ref{fig:mbrg} similarly show Pa$\beta$ and Br$\gamma$ features simultaneously collected with the He~{\scshape i} and Pa$\gamma$ features presented in Figure~\ref{fig:mpag}.  Pa$\beta$ has a larger flux than Pa$\gamma$ and, as a result, is more easily detected during the quiescent accretion phase of the orbit.  The moderate resolution spectra provide insight into the structure of the line and a higher signal-to-noise ratio with which to better estimate the mass accretion rate.  In most epochs for which we have moderate resolution spectra, the feature appears slightly asymmetric with a sharp edge on the red side of the line.  Due to the presence of a notable photospheric absorption feature \citep[unidentified by][]{rayn2009} in the spectra of late K and early M main-sequence stars, we conclude that the asymmetry of Pa$\beta$ is most likely due to this photospheric feature.  For all epochs, there is no clear evidence in any of the Paschen and Brackett series features for the red-shifted absorption component frequently reported for hydrogen lines in the near-IR spectra of accreting TTSs.

\subsection{Quiescent Phase Mass Accretion Rates}

\begin{figure}[ht]
\begin{center}
	\includegraphics[angle=0,width=0.9\columnwidth]{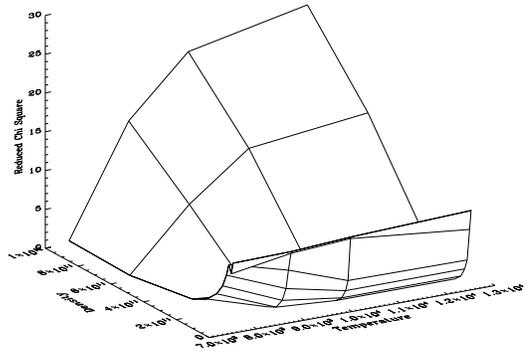}
	\caption{Plotted as a two-dimensional surface are the reduced $\chi^2$-values for the different temperature and density model fits to the observed Paschen decrement values from 2003 December 04 UT.  The ``trough'' in the front of the surface locates the models that are the best fits to the data and indicates the small range of best fit densities, while doing little to constrain the temperatures over the 7500~K to 12500~K range in the models.  For instance, the steep ride of the surface for high temperatures and densities, strongly rule out the possibility of the emitting gas being both hot and dense by KF model standards.}
	\label{fig:surf_plots_a}
\end{center}
\end{figure}

\begin{figure}[ht]
\begin{center}
	\includegraphics[angle=0,width=0.9\columnwidth]{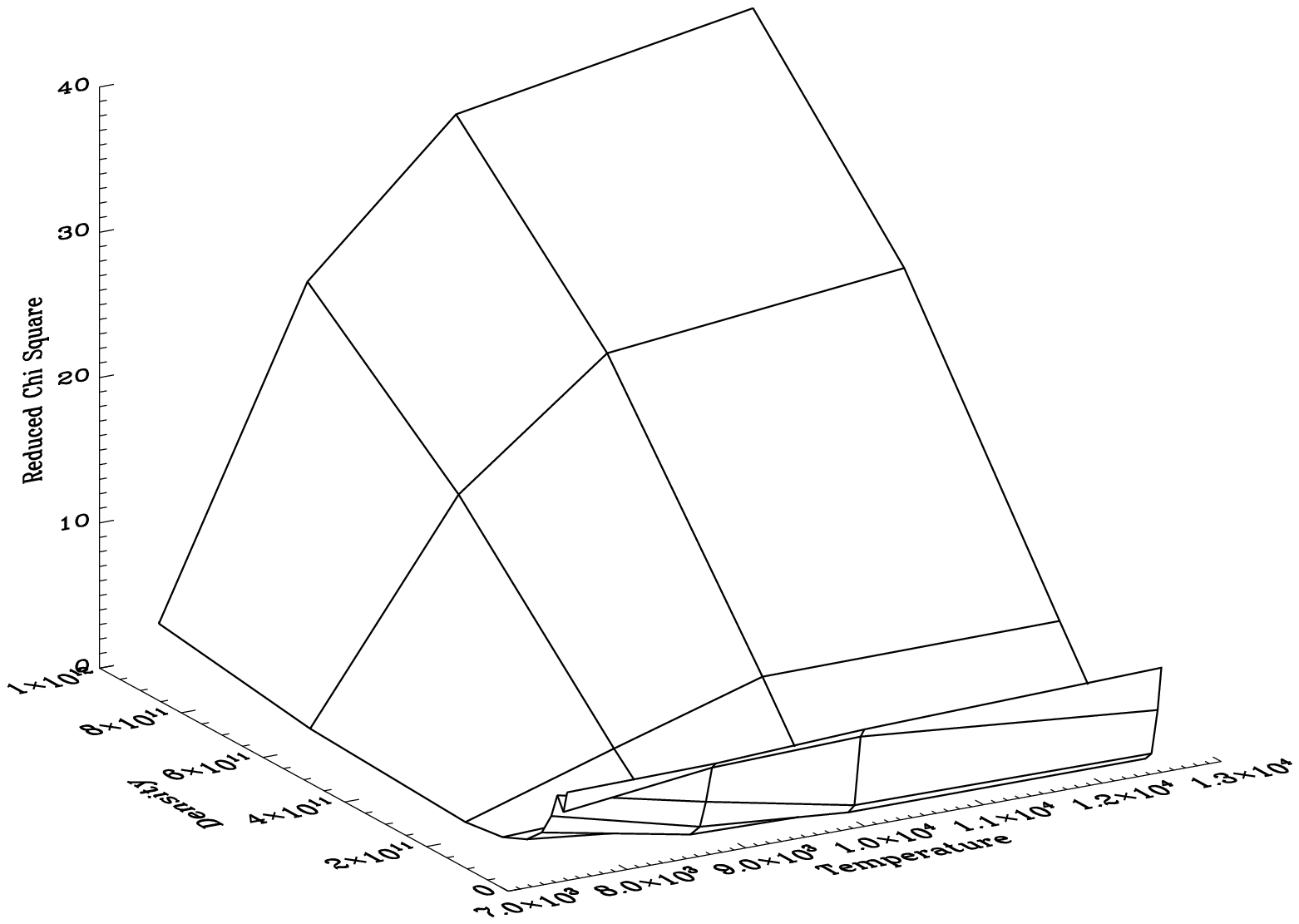}
	\caption{Same as Figure~\ref{fig:surf_plots_a} for observations of DQ~Tau on 2006 October 05 UT.}
	\label{fig:surf_plots_b}
\end{center}
\end{figure}

\begin{figure}[ht]
\begin{center}
	\includegraphics[angle=0,width=0.9\columnwidth]{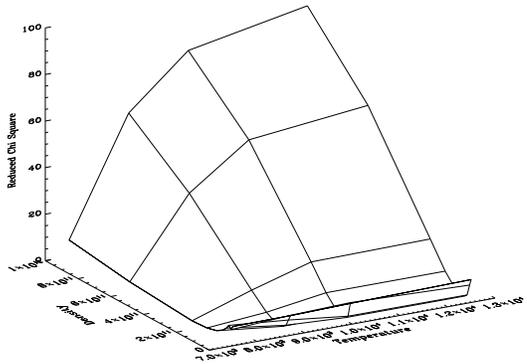}
	\caption{Same as Figure~\ref{fig:surf_plots_a} for observations of DQ~Tau on 2012 October 02 UT.}
	\label{fig:surf_plots_c}
\end{center}
\end{figure}

Mass accretion rates in T Tauri systems are most reliably determined by measuring excess Balmer continuum emission at wavelengths shorter than the Balmer jump \citep{vale1993,gull1998,herc2008}.  However, the observed correlations between mass accretion rates and infrared H~{\scshape i} and the Ca~{\scshape ii} infrared triplet emission line luminosities have led to empirical relationships relating infrared line strengths to mass accretion rates \citep{muze1998a}.  Using the following empirical relations,

\begin{equation}
\log(\frac{L_{acc}}{L_{\odot}}) = (1.14 \pm 0.16)\log(\frac{L_{Pa\beta}}{L_\odot}) + (3.15 \pm 0.58)
\end{equation}

\begin{equation}
\log(\frac{L_{acc}}{L_{\odot}}) = (1.26 \pm 0.19)\log(\frac{L_{Br\gamma}}{L_\odot}) + (4.43 \pm 0.79)
\end{equation}

\noindent
we can calculate the accretion luminosities from the Pa$\beta$ and Br$\gamma$ line luminosities.  We then use the following formula,

\begin{equation}
\log\left(\frac{\dot{M}}{M_\odot}\right) = \log\left(\frac{L_{acc}}{L_\odot}\right)  +\log\left(\frac{R_{\ast}}{R_\odot}\right)  -\log\left(\frac{M_{\ast}}{M_\odot}\right) - \log\left(-7.18\right)
\end{equation}

\noindent
to convert the accretion luminosity to a mass accretion rate.  In Table~2, we present the equivalent widths and upper limits of the four strongest hydrogen emission lines (Pa$\beta$, Pa$\gamma$, Pa$\delta$, and Br$\gamma$) and the 1.083~$\mu$m He~{\scshape i} feature as well as the accretion luminosities and mass accretion rates.  We note that the $\log\frac{\dot{M}}{M_\odot}$ of the accretion luminosities presented were determined using the stronger, higher signal-to-noise ratio Pa$\beta$ line luminosity.  In Section \S3.6, we discuss the determination of the spectral type of the DQ~Tau system from which we base a mass and stellar radius ($M_\ast$~=~1.3~M$_\odot$ and $R_\ast$~=~1.6~R$_\odot$, respectively) for the determination of $\dot{M}$.  In Figure \ref{fig:lacc}, we plot the accretion luminosity versus orbital phase.  
%tFor the epochs of observations that occur outside of the periastron passages, we estimate a value of 1.28$\pm^{0.34}_{0.26}$~$\times$~10$^{-9}$~M$_\odot$~yr$^{-1}$ for the average quiescent mass accretion rate, which serves as useful comparison of accretion activity at the periastron flares and the one anomalous flare observed near apastron.  We note that our value for the quiescent mass accretion rate compares favorably with the estimate of 6~$\times$~10$^{-10}$~M$_\odot$~yr$^{-1}$ \cite{gull1998}, which was derived from the excess Balmer continuum emission.

The randomly collected data sample the strengths of the emission features throughout the entire orbital phase, though they do not provide continuous coverage of any one full orbit of the system.  Using H~{\scshape i} and the Ca~{\scshape ii} lines as indicators of accretion activity, we detect evidence of accretion flares in eight epochs of observations located near four different periastron passages ($\phi$~$=$~0.033, 0.037, 0.713, 0.876, 0.897, 0.961, 0.962, and 0.975; see Table ~\ref{tbl:2}).  The flares are denoted by a strengthening of several Paschen and Brackett series H~{\scshape i} emission features, as is shown in Figures~\ref{fig:mpag}, \ref{fig:mpab}, and \ref{fig:mbrg}.   Upper limits are placed on the strengths of the emission features during the quiescent portion of the orbit when the line strengths fall below our sensitivity thresholds (see Figure~\ref{fig:lacc}). To determine a quiescent (non-flaring) accretion rate, we select epochs which do not indicate flaring activity (weak Hydrogen features, low intra-night variability) to estimate the mean quiescent mass accretion rate as 6.72~$\pm$~$^{3.09}_{2.17}$~$\times$~10$^{-10}$~M$_\odot$~yr$^{-1}$ (this corresponds, via linear mapping, to a quiescent accretion luminosity of -1.76~$\pm$~0.1~$\log\left(\frac{L_{acc}}{L_\odot}\right)$). We note that our value for the quiescent mass accretion rate compares favorably with the estimate of 6~$\times$~10$^{-10}$~M$_\odot$~yr$^{-1}$ \citep{gull1998}, which was derived from the excess Balmer continuum emission.

\begin{deluxetable*}{lccccccc}
\tabletypesize{\tiny}
\tablecaption{DQ~Tau Line Measurements \label{tbl:2}}
\tablewidth{0pt}
\tablehead{
				%\colhead{UT Date}
			%&
			  	\colhead{Orbital}
			&  	\colhead{He~{\scshape i}}
			&  	\colhead{Pa$\delta$}
			&      \colhead{Pa$\gamma$}
			&	\colhead{Pa$\beta$}
			&	\colhead{Br$\gamma$}
			&	\colhead{$\log\left(\frac{L_{acc}}{L_\odot}\right)$}
			&	\colhead{$\log\left(\frac{\dot{M}}{M_\odot {\rm yr}^{-1}}\right)$}  \\
				%\colhead{} 
			%&
				\colhead{Phase\tablenotemark{a}}
			&	\colhead{W{$_\lambda$}(\AA)}
			&	\colhead{W{$_\lambda$}(\AA)}
			&	\colhead{W{$_\lambda$}(\AA)}
			&	\colhead{W{$_\lambda$}(\AA)}
			&	\colhead{W{$_\lambda$}(\AA)}
			&	\colhead{}
			&	\colhead{}}
\startdata
%2005/10/16    &  
0.033 $\pm$ $^{0.031}_{0.093}$ & -11.40$\pm$0.52 & -5.19$\pm$0.52&-6.59$\pm$0.52&-8.91$\pm$0.62&-6.63$\pm$1.03&-1.03$\pm$0.05&-8.44$\pm$0.05\\
%2006/01/19    & 
0.037 $\pm$  $^{0.034}_{0.102}$ & -28.67$\pm$0.52  &-6.19$\pm$0.52 &-11.05$\pm$0.52&-16.53$\pm$0.62&-9.6$\pm$1.03&-0.72$\pm$0.03&-8.13$\pm$0.03\\
%2006/10/04    & 
0.372 $\pm$ $^{0.035}_{0.107}$  & -8.57$\pm$0.52    &-1.01$\pm$0.52&-4.51$\pm$0.52&-3.27$\pm$0.62&\nodata&-1.53$\pm$0.16&-8.94$\pm$0.16\\
%2011/11/12    & 
0.402 $\pm$ $^{0.048}_{0.139}$  & -11.40$\pm$0.19 & -0.70$\pm$0.19 & -1.77$\pm$0.19 &-2.30$\pm$0.23 & -1.462$\pm$0.38 &-1.70$\pm$0.07&-9.11$\pm$0.07\\
%2006/10/05   & 
0.433 $\pm$  $^{0.034}_{0.102}$ &-8.89$\pm$0.52     &-2.22$\pm$0.52&-2.98$\pm$0.52&-6.16$\pm$0.62&-3.49$\pm$1.03&-1.21$\pm$0.08&-8.62$\pm$0.08\\
%2003/12/13   &  
0.437 $\pm$ $^{0.028}_{0.084}$  & -8.84$\pm$0.52     &\nodata&\nodata&-1.78$\pm$0.62&\nodata &-1.83$\pm$0.35&-9.24$\pm$0.35\\
%2011/11/13   & 
0.466 $\pm$ $^{0.048}_{0.139}$  &-10.45$\pm$0.19  &-0.85$\pm$0.19&-2.29$\pm$0.19&-2.36$\pm$0.23&-1.55$\pm$0.38&-1.69$\pm$0.07&-9.10$\pm$0.07\\
%2005/01/29   & 
0.479 $\pm$  $^{0.031}_{0.093}$ &-6.77$\pm$0.52      &-2.70$\pm$0.52&-3.26$\pm$0.52&-2.40$\pm$0.62&-2.43$\pm$1.03&-1.68$\pm$0.23&-9.09$\pm$0.23 \\
%2003/12/14   & 
0.500 $\pm$ $^{0.028}_{0.084}$   & -7.61$\pm$0.52   &\nodata&\nodata&\nodata&\nodata &-1.82$\pm$0.34&-9.23$\pm$0.34\\
%2011/11/14   & 
0.528 $\pm$ $^{0.048}_{0.139}$  &-10.99$\pm$0.19 &-0.79$\pm$0.19&-1.89$\pm$0.19&-2.95$\pm$0.23&-1.74$\pm$0.38&-1.58$\pm$0.05&-8.99$\pm$0.05\\
%2010/01/04   & 
0.539 $\pm$  $^{0.045}_{0.135}$ & 0.94,-7.80$\pm$0.09& -0.82$\pm$0.09&-1.22$\pm$0.09&-1.80$\pm$0.11&-2.78$\pm$0.18 &-1.82$\pm$0.04&-9.23$\pm$0.04\\
%2010/01/04   & 
0.546 $\pm$  $^{0.045}_{0.135}$ & 1.01,-7.00$\pm$0.09& -0.97$\pm$0.09&-1.16$\pm$0.09& -1.41$\pm$0.11&-1.41$\pm$0.13&-1.94$\pm$0.06&-9.35$\pm$0.06\\
%2003/12/15   & 
0.565 $\pm$ $^{0.028}_{0.084}$  & -7.49$\pm$0.52&\nodata&-1.56$\pm$0.52&-2.43$\pm$0.62&\nodata&-1.67$\pm$0.23&-9.08$\pm$0.23 \\
%2003/12/16   & 
0.627 $\pm$ $^{0.028}_{0.084}$  &-7.66$\pm$0.52&-2.06$\pm$0.52&-3.04$\pm$0.52&-2.07$\pm$0.62&-3.01$\pm$1.03&-1.80$\pm$0.32&-9.21$\pm$0.32 \\
%2010/01/06   & 
0.666 $\pm$ $^{0.045}_{0.135}$  & 1.02,-7.44$\pm$0.09 &\nodata&-1.11$\pm$0.09&-1.65$\pm$ 0.11&-1.61$\pm$0.18&-1.86$\pm$0.05&-9.27$\pm$0.05\\
%2006/01/15   & 
0.713 $\pm$ $^{0.034}_{0.102}$  & -10.90$\pm$0.52 &-1.90$\pm$0.52 &-5.97$\pm$0.52&-5.59$\pm$0.62&\nodata&-1.26$\pm$0.09&-8.67$\pm$0.09\\
%2003/12/04   & 
0.876 $\pm$ $^{0.027}_{0.082}$   & -9.54$\pm$0.52  &-6.24$\pm$0.52&-6.67$\pm$0.52&-11.98$\pm$0.62&-6.75$\pm$1.03&-0.88$\pm$0.04&-8.44$\pm$0.04  \\
%2006/01/18  & 
0.897 $\pm$ $^{0.034}_{0.102}$   & -22.85$\pm$0.52 & -3.86$\pm$0.52 & -4.92$\pm$0.52 &-9.18$\pm$0.62 &-5.71$\pm$1.03&-1.02$\pm$0.05&-8.43$\pm$0.05\\
%2012/10/02 & 
0.961 $\pm$ $^{0.051}_{0.148}$    & 1.56,-21.45$\pm$0.19 & -6.29$\pm$0.19  & -12.56$\pm$0.19 &-16.56$\pm$0.23 &-6.25$\pm$0.38&-0.72$\pm$0.01&-8.13$\pm$0.01  \\
%2012/10/02 & 
0.962 $\pm$ $^{0.051}_{0.148}$ & 1.58,-21.62$\pm$0.19 & -7.90$\pm$0.19  & -12.65$\pm$0.19 & -16.05$\pm$0.23 &-6.17$\pm$0.38&-0.74$\pm$0.01&-8.15$\pm$0.01  \\
%2012/10/02 & 
0.975 $\pm$ $^{0.051}_{0.148}$ & 1.21,-18.52$\pm$0.19 & -5.62$\pm$0.19  & -10.49$\pm$0.19 & -13.80$\pm$0.23 &-4.93$\pm$0.38 &-0.81$\pm$0.01&-8.22$\pm$0.01 \\
\enddata

\tablenotetext{a}{Orbital phases were calculated beginning on Julian date 2449582.54 using the orbital period of 15.8016$\pm$$^{0.002}_{0.006}$ measured by \citet{huer2005}.}
%% Text for table notes should follow after the \enddata but before
%% the \end{deluxetable}. Make sure there is at least one \tablenotemark
%% in the table for each \tablenotetext.
\end{deluxetable*}

\subsection{Pulsed-Accretion Activity: Anomalous Flare Near Apastron}

\begin{figure*}[ht]
\begin{center}
\includegraphics[angle=0,width=1.8\columnwidth]{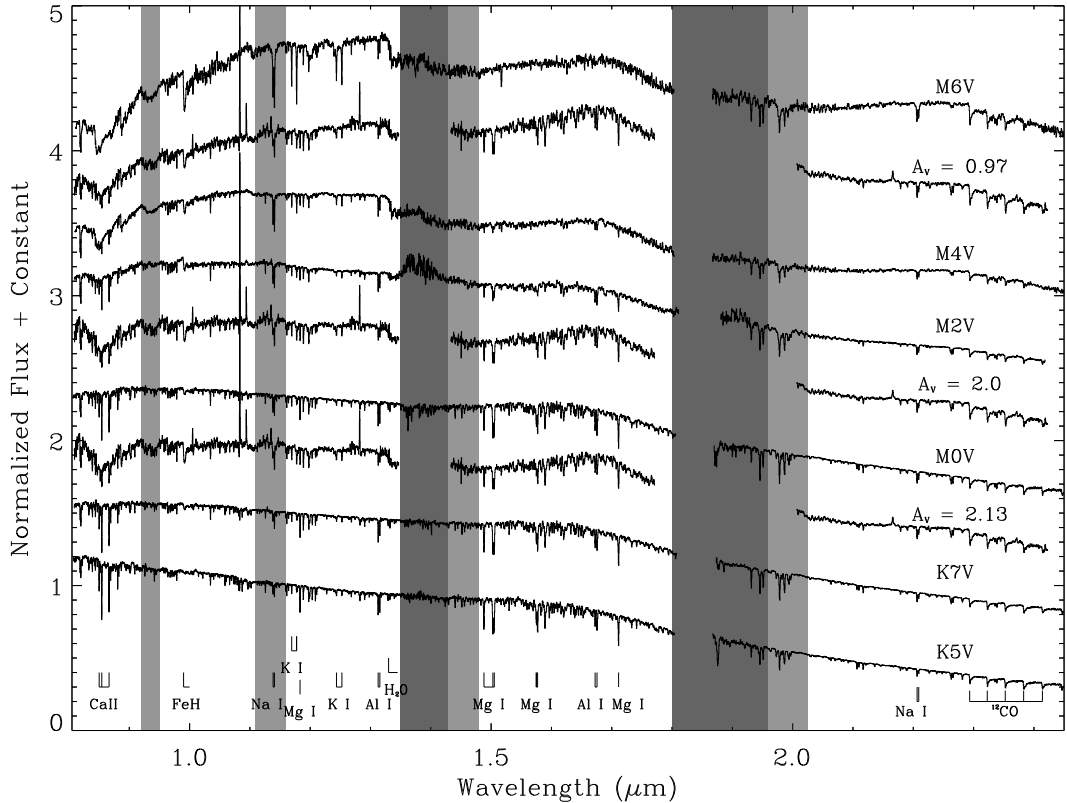}
\caption{A sequence of near-infrared spectra from the Spex IRTF Spectral Library \citep{rayn2009} spanning 0.8 to 2.45~$\mu$m are presented in normalized flux units.  Interspersed is a SpeX observation of DQ~Tau taken on 2011 November 13 UT corrected with the three different published values of the visual extinction.  The three dereddened DQ~Tau spectra are positioned so as to be bracketed by the spectral standards that best match the overall shape of the spectrum.  Many of the strongest metallic photospheric absorption features are identified.  Following the plots from \citet{vacc2011}, regions of poor atmospheric transmission ($<$20\%) are denoted by dark gray and regions of moderate atmospheric transmission ($<$80\% are denoted by light gray vertical bars.  }
\label{fig:fullspec}
\end{center}
\end{figure*}

\begin{figure*}[ht]
\begin{center}
\includegraphics[angle=0,width=1.8\columnwidth]{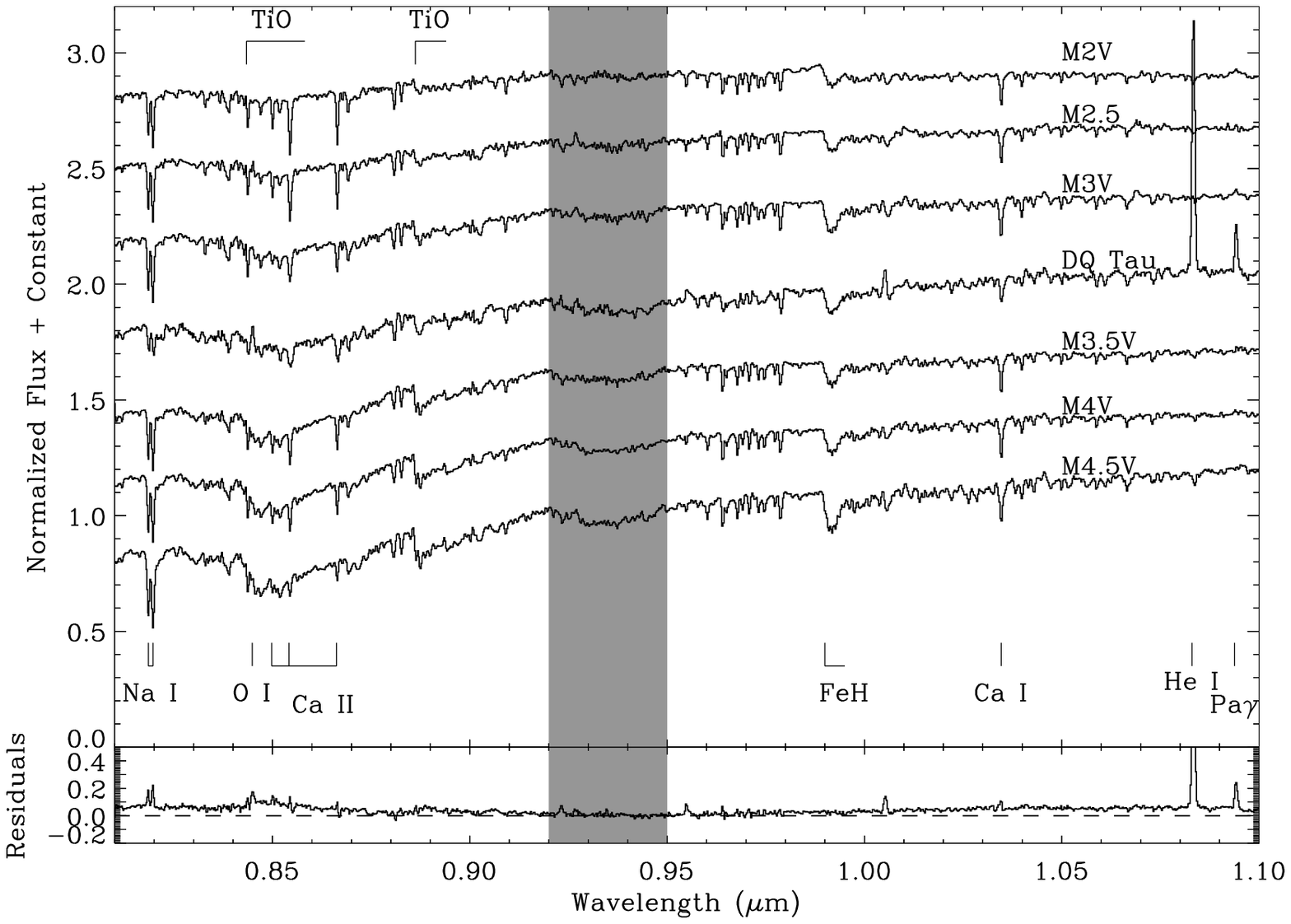}
\caption{A comparison of M dwarf spectra from the SpeX library to one epoch of DQ~Tau taken on 2011 November 14 UT over 0.8-1.1~$\mu$m.  All spectra have been dereddened and normalized.  A visual extinction of $A_V$~=~0.97 \citep{keny1995} was used for DQ~Tau.  The spectral sequence of M dwarfs chosen to bracket the DQ~Tau spectra is based primarily on the strengths of the TiO bands at 0.85 and 0.88~$\mu$m and the FeH band at 0.99~$\mu$m.  The residual spectrum is the difference between the DQ~Tau and the M3.5V spectrum.}
\label{fig:pt8}
\end{center}
\end{figure*}

Selecting the previously-defined quiescent observations, we find (as noted above) a mean accretion luminosity of $\log\left(\frac{L_{acc}}{L_\odot}\right) = -1.76$, with a dispersion of $\sigma=0.1$. Interpreted in this context, the first night of the proposed anomalous flare at $\phi=0.372$ is a 2.3-$\sigma$ outlier, while the second night of the flare at $\phi=0.433$ is a 5.51-$\sigma$ outlier. In fact, the line strengths and accretion luminosities on these two nights compare more favorably to the periastron accretion flares, which have a mean accretion luminosity of $\log\left(\frac{L_{acc}}{L_\odot}\right) = -0.845$ and a dispersion of $\sigma=0.125$. The second night of the flare is more consistent with originating in the periastron flaring distribution (3.65-$\sigma$) than the quiescent distribution (5.51-$\sigma$). Note also that an observation at $\phi=0.71$ exhibits a 5.01-$\sigma$ flare.  Based on the photometric monitoring of \citet{math1997}, such a flare constrained to within $\Delta\phi$~=~$\pm$~0.3 of periastron, would have been considered one of the periastron flares.  Therefore, we did not consider it part of the quiescent portion of the orbit nor anomalous in anyway.  Unfortunately, additional observations from the same orbit were not acquired and we cannot comment on whether the system is simply experiencing an early beginning to a periastron accretion flare or if this is a completely random event.  The data also do not posses temporal coverage within 0.8~$\le$~$\phi$~$\le$~0.35 to definitively say that infrared periastron flares are similarly constrained in phase as the optical flares.

The fact that the anomalous apastron flare or outburst occurs when the stars are near their greatest separation strengthens the conclusions of \citet{jens2007} that the flaring events are not solely produced by the interaction of the stellar magnetospheres, a potential flaring mechanism originally proposed by \citet{basr1997}.  Neither the hydrodynamic model of \citet{arty1996} nor that of \citet{gunt2002} predict significant pulsed accretion activity occurring near periastron.  Unfortunately, we only have two observations preceding the apastron passage.  As a result, we do not know if an increase in the accretion activity continues as the binary approaches apastron and whether that behavior is then followed by a decay or if the increased activity continues through the rest of the orbit.  Previous data collected at multiple wavelengths show relatively little variation in the quiescent emission levels of the H~{\scshape i} features at both optical and infrared wavelengths.

The simulations of \citet{gunt2002} and \citet{deva2011} provide a framework for considering the nature of the accretion flare observed at apastron.  Both simulations predict that perturbations in the circumbinary disk will lead to the formation of irregularly-shaped, non-axisymmetric structures extending inwards from the inner edge of the disk.  The presence of such structures coinciding with the close passage of one or both stellar cores at apastron, may lead to an enhancement of the mass accretion rate such as that observed in the anomalous flare.  If the size, shape, density, and position of these structures vary considerably over time, such events may be both rare and aperiodic.  The viscosity of the disk is a key parameter to the size and extent of such structures.  Therefore, additional detections of apastron flares, including durations and distances from the disk at which the flares begin should provide rough constraints on the viscosity of material in the circumbinary disk.  We plan to continue the spectroscopic monitoring of DQ~Tau and extend these observations to high-resolution to compare kinematics of the accretion line diagnostics in both the quiescent and flare states.

\subsection{Physical Conditions of the Accreting Hydrogen Gas}

\begin{figure*}[ht]
\begin{center}
\includegraphics[angle=0,width=1.8\columnwidth]{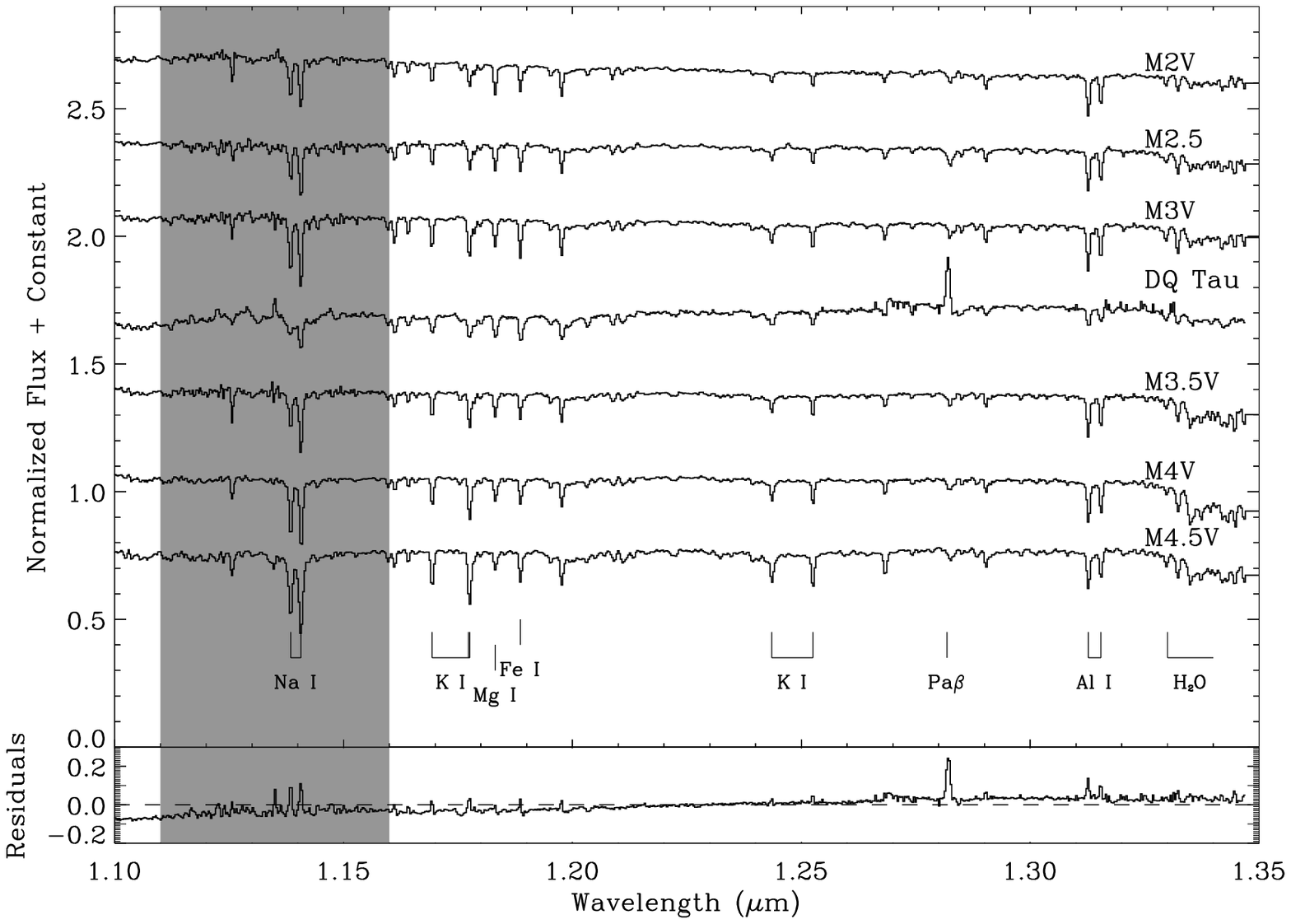}
\caption{A comparison of M dwarf SpeX library spectra to one epoch of DQ~Tau taken on 2011 November 13 UT over 1.1-1.35~$\mu$m.  The spectral sequence of M dwarfs is chosen to bracket the DQ~Tau spectra based primarily on the strength of the H$_2$O feature at 1.33~$\mu$m.}
\label{fig:1pt1}
\end{center}
\end{figure*}

\begin{figure*}[ht]
\begin{center}
\includegraphics[angle=0,width=1.8\columnwidth]{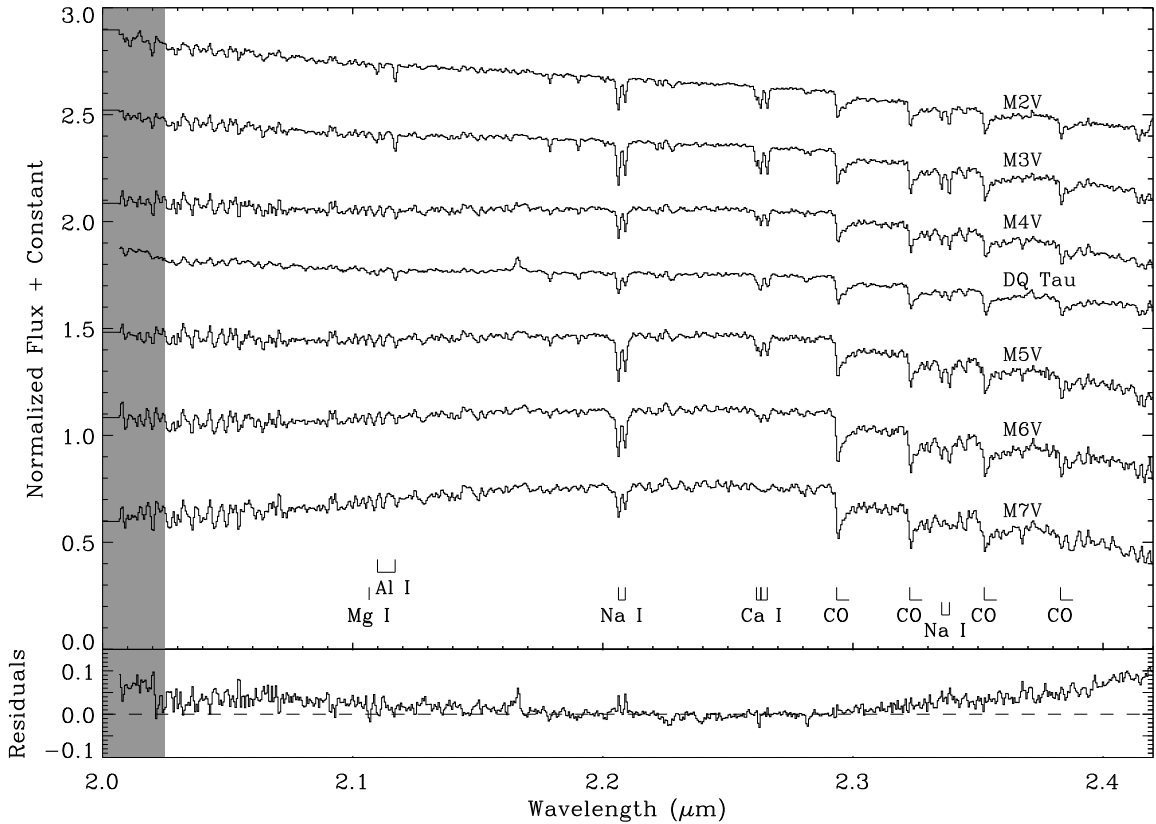}
\caption{A comparison of M dwarf SpeX library spectra to one epoch of DQ~Tau taken on 2011 November 13 UT over 2.0-2.4~$\mu$m.   Note that the strengths of the CO bandheads at 2.2935, 2.3227, 2.3525, and 2.3830~$\mu$m are more similar to those of the later M5-6V than the earlier M2-4V spectral types.  However, the strengths of the Na~{\scshape i}, Mg~{\scshape i}, Al~{\scshape i}, and Ca~{\scshape i} metallic lines more closely agree with the earlier spectral types.}
\label{fig:2pt0}
\end{center}
\end{figure*}

\begin{figure*}[ht]
\begin{center}
\includegraphics[angle=0,width=1.8\columnwidth]{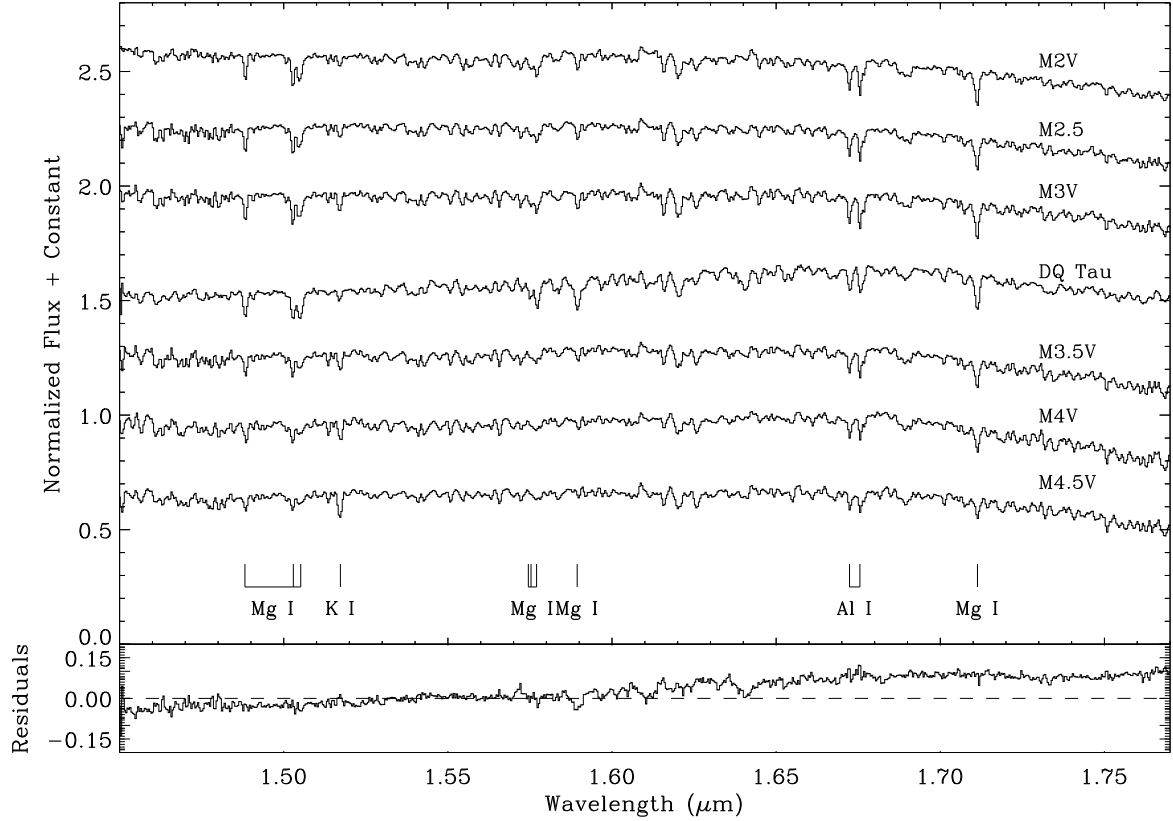}
\caption{A comparison of M dwarf spectra from the SpeX library to one epoch of DQ~Tau taken on 2011 November 13 UT over 1.45-1.77~$\mu$m.  The same M dwarf spectral standards presented in Figure~\ref{fig:pt8} are presented here over a region of the spectrum containing several temperature sensitive metallic absorption features.  Note that the strengths of the Mg~{\scshape i}, Al~{\scshape i}, and K~{\scshape i} features clearly suggest a spectral type earlier than an M2V for DQ~Tau.}
\label{fig:1pt45um}
\end{center}
\end{figure*}

\begin{figure*}[ht]
\begin{center}
\includegraphics[angle=0,width=1.8\columnwidth]{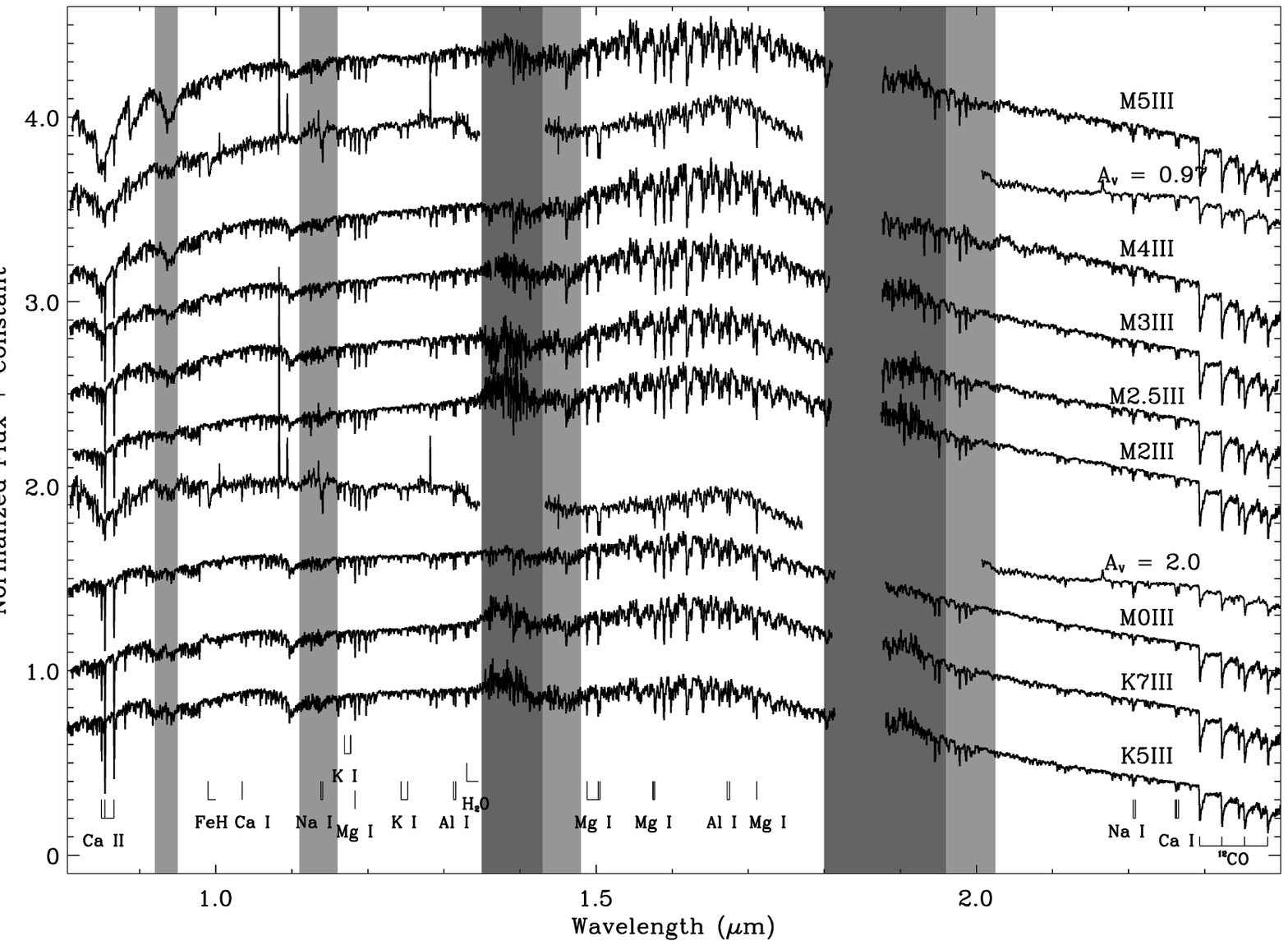}
\caption{A comparison of M and K giant star standard spectra to DQ~Tau spectra (2011 November 13 UT)  dereddened with two values of interstellar extinction A$_v$ = 0.97 and 2.0.  The TiO feature at 0.85~$\mu$m is considerably stronger in the DQ~Tau spectrum than that of the early M and late K giant spectral standards.  Also note the absence and relative weakness of the FeH feature at 0.99~$\mu$m in the giant star spectra.}
\label{fig:giants}
\end{center}
\end{figure*}

In addition to being excellent proxies for measuring the mass accretion rates in TTSs, the multiple near-IR hydrogen lines collected simultaneously in the cross-dispersed data allow us to probe the temperature and density of the accreting gas.  By comparing the values of multiple hydrogen line ratios to those predicted by temperature and density dependent hydrogen line excitation models, we can search for the models that produce the most statistically significant fits to the data and constrain the gas temperature and density.  If the emitting gas is located in the accretion column, either at the point where it is loaded into the flow or where it resides in the column (or both), the physical conditions of this gas may be revealed by such a fitting procedure.  Previously, \citet{bary2008} used line ratios from over 70 spectra of actively accreting TTSs and the well-known case~B line excitation model, which assumes a recombining hydrogen gas in which transitions above the Lyman series are optically thin, to solve for the average temperature and density of accreting gas in TTS systems.  \citet{bary2008} found a range of electron densities (10$^9$~cm$^{-3}$~$\le$~$\rho$~$\le$~10$^{11}$~cm$^{-3}$) that were in good agreement with those predicted by magnetospheric accretion models.  However, the temperatures constrained by the best fit models ($T_{gas}$~$\le$~2000~K) were considerably lower than the typical 6000~K~$\le$~$T_{gas}$~$\le$~12000~K predicted for accreting gas \citep{muze2001}.  A new line excitation model developed by \citet{kwan2011}, specified for infalling/outflowing gas ($\lvert$$v$$\rvert$~$=$~150~km~s$^{-1}$), includes line opacity as a free parameter, in addition to gas density, kinetic temperature, an ionizing flux, and a local velocity gradient.  The Kwan-Fischer (KF) line excitation model places the emitting gas within a few stellar radii of the stellar surface and includes photoionization as an excitation mechanism.  Using the data from \citet{bary2008}, the KF models found a best fit density of $N_H$~$\sim$~10$^{11}$~cm$^{-3}$ for temperatures in the range of 8750~K~$\le$~$T_{gas}$~$\le$~3$\times$10$^{4}$~K and $N_H$~$\sim$~5~$\times$~10$^{11}$ for temperatures $T_{gas}$~$\le$~7500~K.  Temperatures returned by the KF models are in better agreement with the accretion models.  This result suggests that the case~B models are not the appropriate line excitation models for accreting gas and that the H~{\scshape i} line emission arises from collisional excitation, rather than from a recombining gas.

The DQ~Tau system with its periodic accretion flares provides a unique opportunity to search for changes in the physical conditions of the accreting gas with mass accretion rate.   While we have 20 distinct epochs of observations, ten epochs are low-resolution CorMASS spectra that were not sensitive enough to detect multiple H~{\scshape i} features during the quiescent phases of the orbit, making the line ratio analysis described above impossible with this data set.  Using data from one moderate resolution spectrum during periastron and the two low-resolution spectra that caught the system during a flare, one during periastron and another near apastron, we measured the line ratios for five Paschen series lines.  We compared the observed Pa$n$/Pa$\beta$ decrement for each of these epochs to the range of KF models \citep[10$^{9}$~cm$^{-3}$~$\le$~$N_H$~$\le$~10$^{12}$~cm$^{-3}$;][]{edwa2013}.  In Figure~\ref{fig:pantopaba}, \ref{fig:pantopabb}, and \ref{fig:pantopabc}, we present plots of the Paschen decrement for the five lines with measurable fluxes for these three epochs.  We find that our data constrain the densities to be between $N_H$ equals 6.3~$\times$~10$^{10}$~cm$^{-3}$ and 1.0~$\times$~10$^{12}$~cm$^{-3}$, while placing no meaningful constraint on the gas temperatures between 7500~K to 12500~K.  Figures~\ref{fig:surf_plots_a}, \ref{fig:surf_plots_b}, and \ref{fig:surf_plots_c} present three-dimensional surface plots of the reduced $\chi^2$-values for the density and temperatures.  The ``trough" that runs from low temperature to high temperature over a small range of densities clearly illustrates the wide range of acceptable temperatures for the accreting gas, while simultaneously demonstrating the KF models ability to constrain the gas density.  Given the large uncertainties in our measurements of the line ratios, we cannot constrain the range of temperatures and densities well enough over the three epochs to detect variations of the physical conditions of the accreting gas.  In the future, with the addition of more high signal-to-noise resolution spectra during both quiescent and accretion flare epochs, we hope to search for variations in the physical conditions of the gas as a function of accretion activity. 

\subsection{Spectral Classification}

The moderate-resolution near-IR spectra collected with SpeX and TSpec provide an opportunity to measure a spectral type for DQ~Tau from a wavelength region inaccessible to previous spectroscopic studies that were centered only on visible wavelengths \citep[e.g.,][]{joy1974,herb1977,basr1997}.  Using the IRTF SpeX Spectral Library \citep{rayn2009}, we begin by following a similar spectral typing procedure outlined in \citet{vacc2011}, which involves comparisons of spectral shapes and the strengths of molecular and metallic features.  We make additional comparisons to giant stars using spectral standards from the SpeX library and model spectra from \citet{coel2005} to test surface gravity effects.  After finding compelling evidence for a much cooler spectral type than previously reported for DQ~Tau and demonstrating the inability of variations in surface gravity to simultaneously account for the observed strengths of the TiO and FeH bands, we investigate the potential effects of large cool spots on the surfaces of the stellar companions.  We show that composite spectra of warm photospheres and large, cool spots produce reasonable fits to the TiO bands at 0.85 and 0.88~$\mu$m and the FeH feature\footnote{The iron hydride band is commonly referred to as the Wing-Ford band \citep{wing1969}} at 0.99~$\mu$m.  These fits also distinguish between the three different values of $A_V$ published for DQ~Tau.  

\subsubsection{IR Spectral Typing}

\begin{figure*}[ht]
\begin{center}
\includegraphics[angle=0,width=1.8\columnwidth]{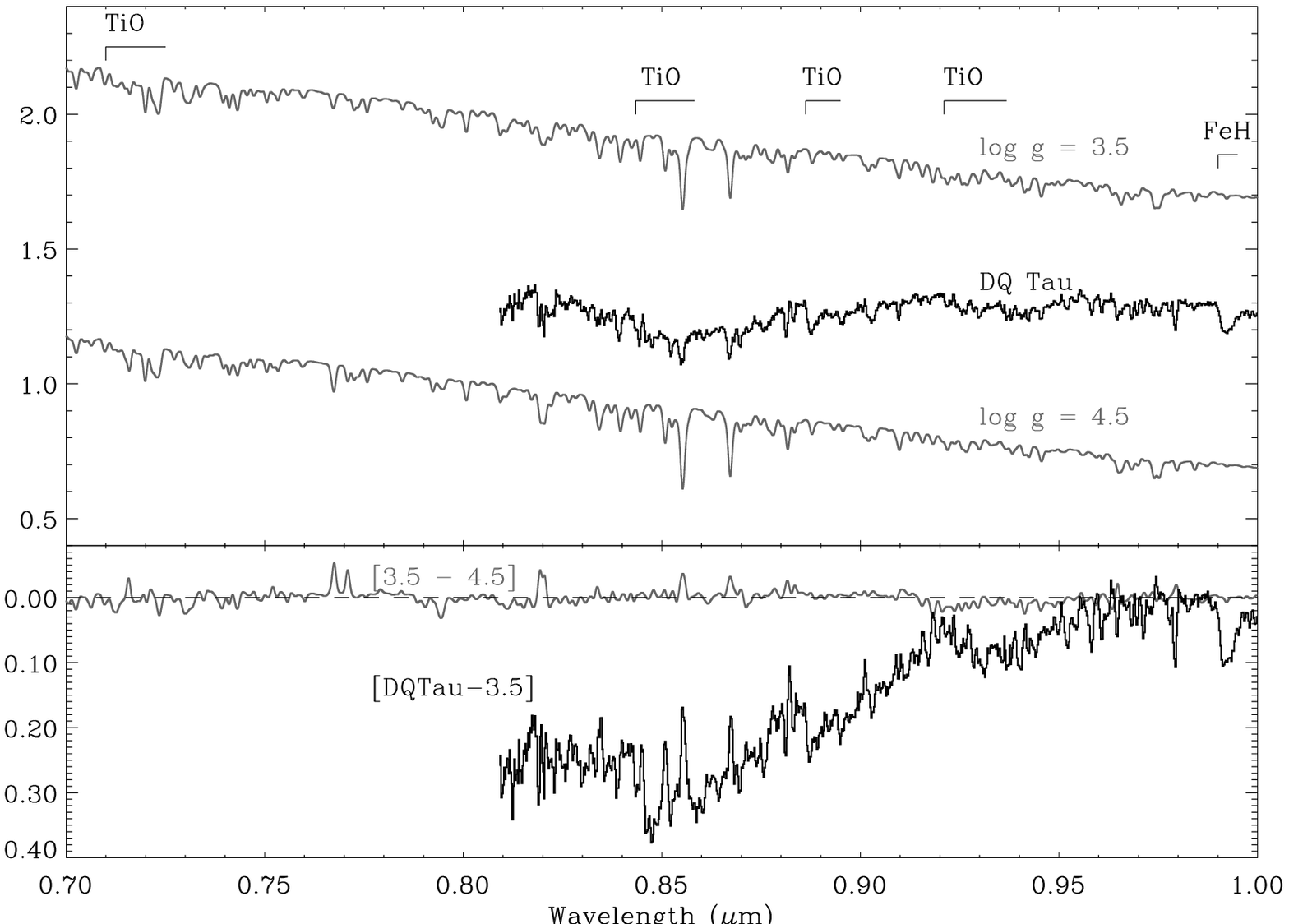}
\caption{In the upper panel, two synthetic spectra (gray) for $T_{eff}$~=~4500~K stars with solar metallicity are plotted with different values for log~$g$, 3.5 and 4.5, from 0.70~$\mu$m to 1.0~$\mu$m includes two TiO bands at 0.71~and 0.85~$\mu$m.  The lower {\it g}-value corresponds to that of a sub-giant, K5IV, while the higher value is that of a K5V star.  Bracketed by the two model spectra is the relevant portion of a dereddened ($A_V$~=~0.97) SpeX spectrum of DQ~Tau (black; 2011 November 13 UT) in which the TiO absorption band centered on 0.85~$\mu$m is clearly visible.  In the lower panel, two residuals are plotted.  The gray residual is the difference of the two synthetic spectra [3.5 $-$ 4.5], while the solid residual is the difference of DQ~Tau and the log~$g$~=~3.5 spectrum.  The small variations in the first residual illustrate that there is little difference in the strength of the TiO band for the difference in log~$g$ between a main sequence and a subgiant star with the same temperature.  The second residual shows that the spectral shape and strength of the TiO feature in DQ~Tau is not approximated well by a T~$=$~4500~K star with either values of log~$g$.}
\label{fig:tio_g}
\end{center}
\end{figure*}

\begin{figure*}[ht]
\begin{center}
\includegraphics[angle=0,width=1.8\columnwidth]{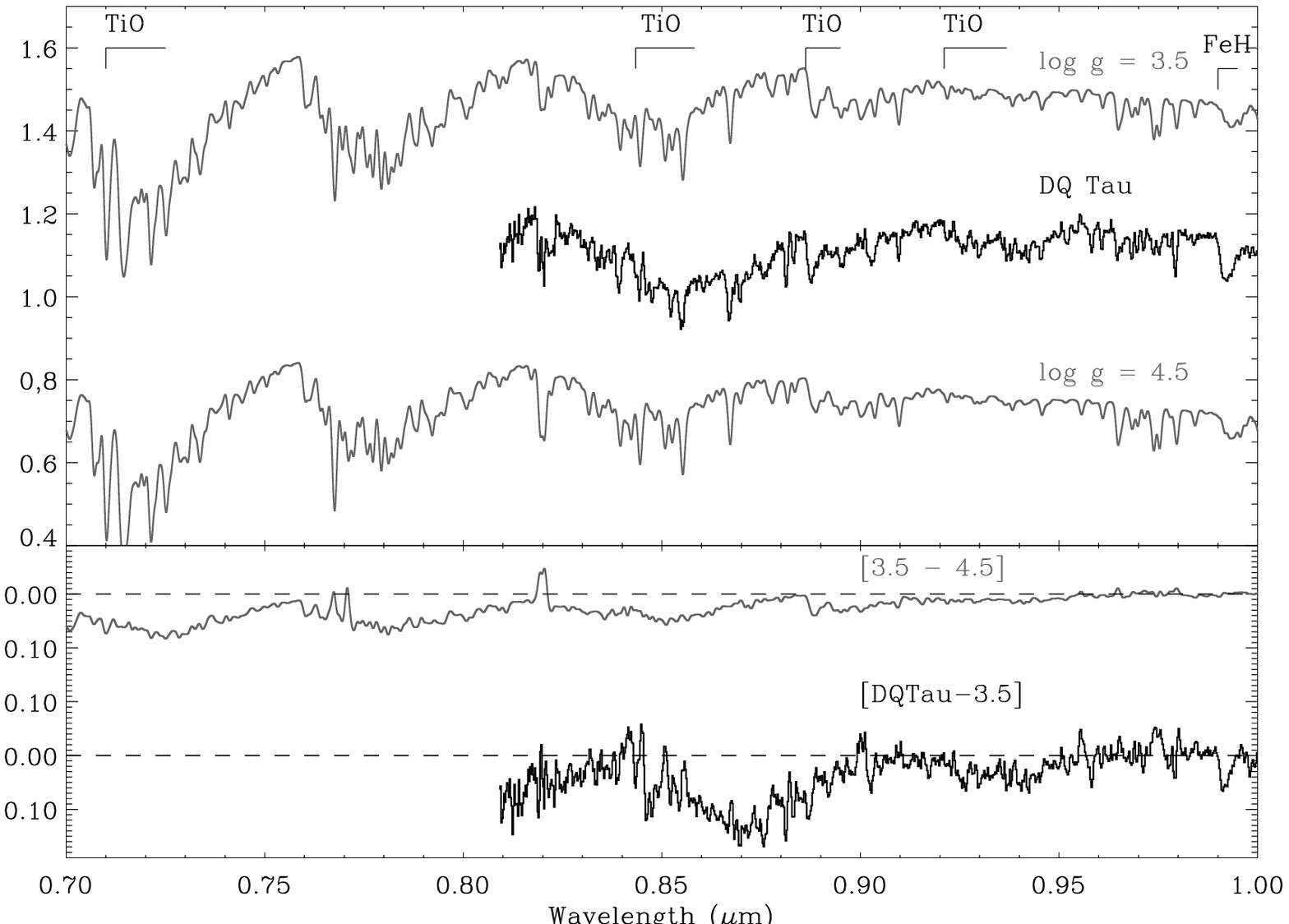}
\caption{Similar to Figure~\ref{fig:tio_g}, in the upper panel, two synthetic spectra for $T_{eff}$~=~3500~K stars (similar to M2) with solar metallicity and log~$g$ of 3.5 and 4.5 bracket a dereddened spectrum of DQ~Tau (2011 November 13 UT).  In the lower panel, the light gray residual is the  difference between the two synthetic spectra [3.5 $-$ 4.5], while the solid line results from subtracting the log~$g$~=~3.5 synthetic spectrum from DQ~Tau.  Note that the spectral shape is fit well with a significant residual left for the longer wavelength portion of the TiO band.}
\label{fig:tio_m2}
\end{center}
\end{figure*}

In Figure~\ref{fig:fullspec}, we present a sequence of spectral standards taken from the SpeX library that spans the range of spectral types previously determined for the DQ~Tau system \citep[M0V, M1V, and K4-5V/M1-1.5V;][]{joy1974,herb1977,basr1997}.  Bracketed by the main-sequence standards that share the same general spectral shape are three DQ~Tau spectra generated from one SpeX observation collected on 2011 November 13 UT and corrected for reddening using three values of the visual extinction ($A_V$) taken from the literature: 0.97, 2.0, and 2.13 \citep{keny1995,math1997,stro1989} and the extinction law found in \citet{mart1990}.  Figure~\ref{fig:fullspec} clearly shows that one magnitude difference in visual extinction ($\Delta$~$A_V$~=~1) can account for considerable variation in the shape of the DQ~Tau spectrum shortward of 1.5~$\mu$m, which would result in a spectral type mismatch.  For example, M4V-M6V seems appropriate for DQ~Tau with an $A_V$ of 0.97 whereas M0V-M2V is best for an $A_V$ of 2.0 or 2.13.

The presence of several molecular absorption features provide another means for spectral typing.  Broad water absorption bands at 1.4 and 1.9~$\mu$m strongly affect the overall spectral shape, producing the hump-like or triangular features in the {\it H}- and {\it K}-bands (see Figure~\ref{fig:fullspec}).  The magnitude of the hump in the {\it H }band is best approximated by the M5V and M6V standards over the entire range of $A_V$, while the feature in the {\it K}-band is better fit by M3-3.5V.  It is important to note that the {\it H}-band spectra of young stars has a more pronounced triangular shape as compared to the more smoothly rounded features in the spectra of evolved field stars \citep{luca2001,aller2007,aller2013}.  Thus the shape of the DQ~Tau spectrum in the {\it H}~band will not be perfectly matched by the spectral shape of the more evolved standards in the SpeX library.  Nonetheless, clear variations in the strength of the {\it H-}band feature are evident in Figure~\ref{fig:fullspec}  { \footnote {We note that the {\it H}-band feature is poorly fit by the M7V spectral standard, making M6V a firm limit to later-type M stars and explaining why it is not included in the figure.}}.

In addition to the broad water absorption features in the DQ~Tau spectrum, we note the presence of the following molecular features: (1) TiO absorption bands at 0.85 and 0.88~$\mu$m, (2) FeH at 0.99~$\mu$m, (3) a strong H$_2$O band at 1.33~$\mu$m, and (4) $^{12}$CO bandheads in the {\it K}-band.  The strengths of the TiO absorption bands at 0.85 and 0.88~$\mu$m, though slightly affected by the choice of $A_V$, are much deeper than those of the K5-M2V spectra for all values of $A_V$.  Figure~\ref{fig:pt8} provides a direct comparison between a dereddened DQ~Tau spectrum (2011 November 14 UT\footnote {A different night is used for this figure due to the high quality of data near 0.8~$\mu$m.  We note that the accretion activity was measured to be the same on both nights.}; $A_V$~=~0.97) and the spectral standards that bracket DQ~Tau between 0.8 and 1.1~$\mu$m.  Based on the strengths of the TiO bands and the FeH feature, fits between the main-sequence standards and DQ~Tau constrain the spectral type to be between an M3 and M4V over the 0.85-1.0~$\mu$m wavelength range, a value 2-3 spectral classes cooler than quoted in the literature.  \citet{schi1997} clearly demonstrate that the FeH feature heavily depends on the surface gravity with cool dwarf stars possessing the strongest FeH features.  Therefore, the strength of the FeH feature, in addition to suggesting a cooler spectral type, also indicates a {\it dwarf-like} surface gravity for these stars.

Figure~\ref{fig:1pt1} shows the DQ~Tau spectrum between 1.1 and 1.35~$\mu$m, which includes the H$_2$O feature at 1.33~$\mu$m.  The strength of the H$_2$O feature is best fit in the M2.5V to M3.5V spectral type range.  Figure~\ref{fig:2pt0} contains the {\it K}-band and several CO bandhead absorption features.  The CO feature strengths were best fit between an M3.5V and M4.5V, in rough agreement with the spectral type suggested by the spectral shape in the {\it K}-band.  Any significant continuum veiling in the {\it K}-band due to an infrared excess would result in a determination of spectral types corresponding to {\it higher} temperatures than the that of the actual photosphere.  Therefore, M3.5V-M4.5V are upper limits to the spectral type determination from the {\it K}-band.  Regardless of the visual extinction value we choose to adopt, the strengths of these {\it molecular} features collectively and definitively suggest a spectral type cooler than those previously proposed for the DQ~Tau system, with the earliest type being an M2.5V and other features suggesting a type as cool as an M6V. 

On the other hand, the strength of temperature sensitive metal lines such as K~{\scshape i} at 1.16934, 1.17761, and 1.51725~$\mu$m (see Figures~\ref{fig:fullspec} and \ref{fig:1pt1}) and several Mg~{\scshape i} features at 1.18314, 1.4881874, 1.50518, 1.57450, 1.57533, and 1.71133~$\mu$m (see Figures~\ref{fig:1pt1} and \ref{fig:1pt45um}) indicate a better match with slightly early-M-type spectral standards, M2V and M0V, in agreement with those found in the literature.  \citet{joy1974} used the strengths of TiO absorption bands at visible wavelengths to determine a spectral type of M0V.  \citet{herb1977} later classified DQ~Tau as an M1V based on ratios of metallic lines in the optical (5850~$<$~$\lambda$~$<$~6700~\AA).  Most recently, \citet{basr1997}, using high-resolution echelle spectra, reported an M1-1.5V type based upon the strength of the TiO band at 7125~\AA.  However, in the same study, a comparison of the ratio of the temperature sensitive 6210.7~\AA\ Sc~{\scshape i} line to a pair of nearby temperature insensitive Fe~{\scshape i} lines resulted in a far earlier spectral type of K4-5V.  Continuum veiling was discussed as a potential explanation for the inconsistency in spectral types \citep{basr1997}.  However, the line ratios were formed from spectral features in a narrow region of the spectrum, which should have mitigated any effects of veiling on the spectral type.  \citet{basr1997} also suggested that the discrepancy may be due to cool star spots or strong chromospheric activity, but suggested that clear evidence in support of either scenario did not exist.   Therefore, Basri et al.\ concluded that the lower surface gravity of the ``puffed-up'' T Tauri star photospheres may lead to a strengthening of the TiO band, resulting in a cooler spectral type relative to that determined by the metallic lines.  

\subsubsection{Testing Spectral-Type Dependence on $\log g$}

The SpeX Spectral Library also includes spectral standards for giant stars over the spectral range presented in Figure \ref{fig:fullspec}; these standards enable us to test the suggestion that the ``puffed-up'' photospheres affect the spectral typing.  In other words, do warm ``puffed-up" or giant stars have TiO bands in the IR with strengths comparable to those observed in DQ~Tau?  In Figure \ref{fig:giants}, we compare two dereddened DQ~Tau spectra to the spectra of giant stars in a spectral type range similar to those presented in Figure \ref{fig:fullspec}.  By visual inspection, the DQ~Tau TiO features at 0.85, 0.86, and 0.88~$\mu$m are clearly stronger than that of the giant stars as cool as an M4III.  The low pressure photospheres of giant stars cannot account for the optical/IR spectral type discrepancy we observe for DQ~Tau.  In addition to the TiO feature, the giant spectra also appear to lack the pronounced triangular spectral shape in the $H$-band and have comparatively weak FeH and H$_2$O absorption features also suggesting that the photospheres in the DQ~Tau system are more like that of a dwarf than a giant star.

We investigated this further, using a library of synthetic spectra (0.3~$\mu$m~$\le$~$\lambda$~$\le$~1.8~$\mu$m) calculated by \citet{coel2005} to probe the dependence of the TiO and FeH feature strengths on surface gravity.  The library covers a broad range of effective temperatures (3500~K~$\le$~T~$\le$~7000~K), metalliticies (-2.5~$\le$~[$\frac{Fe}{H}$]~$\le$~+0.5), and surface gravities (0.0~$\le$~log~$g$~$\le$~5.0).  Figure~\ref{fig:tio_g} plots two model spectra over a wavelength region that includes a portion of the visible and very near-IR (0.65~$\mu$m~$\le$~$\lambda$~$\le$~1.0~$\mu$m) for two stars with T~$=$~4500~K \citep[roughly a K4-5 per][]{basr1997}, solar metallicity, and $\log$~$g$ values of 3.5 and 4.5.  The lower $\log$~$g$ value of 3.5 represents a ``puffed up'' TTS\footnote {Using the stellar parameters from \citet{math1997} of 0.65~M$_\odot$ and 1.6~R$_\odot$ for the mass and radius for both stellar sources in DQ Tau, we calculated a log~$g$ of 3.8.  For the purpose of this comparison plot, we rounded down to 3.5 rather than up to 4.0 to magnify any potential dependence on log~$g$.}, while 4.5 corresponds to a main-sequence star.  The two model spectra were normalized over the same wavelength region and subtracted to produce the residuals plotted in the bottom panel.  The magnitude of the residuals clearly shows that there is no discernible difference in the TiO features at 0.71 and 0.85~$\mu$m between the spectra with different $\log~g$ values.  A similar differencing of spectra between the target DQ~Tau and the $\log~g$~=~3.5 model spectrum ([DQTau - 3.5]) shows that the strength of the TiO and the FeH absorption features in DQ~Tau far exceeds that of both model spectra.  If the DQ~Tau stars have surface temperatures on the order of a K4 or K5 spectral type, lower surface gravity {\it is not a strong enough effect to account for the strength of the TiO features observed in this study and by \citet{basr1997}}.

Since most studies have determined DQ~Tau to have an M0V to M1.5V spectral type, we also compare the strengths of the TiO bands in the synthetic spectra of two 3500~K stars (roughly the $T_{eff}$ for an M2III-V), again with $\log~g$ values of 3.5 and 4.5 in Figure~\ref{fig:tio_m2}.  A comparison to model spectra in Figure~\ref{fig:tio_g} shows that the TiO features for the cooler stars are clearly much stronger and closer approximations of the TiO absorption in DQ~Tau, as should be expected.  The residuals left by differencing the synthetic spectra with different surface gravities and cooler temperatures show evidence for some dependence on surface gravity with the TiO features in the star with the lower $\log~g$ value appearing stronger.  Once again, we subtract the low surface gravity spectrum from DQ~Tau ([DQ~Tau-3.5]) and find that the TiO features are slightly stronger in the photospheres of the DQ~Tau stars.  However, what is most telling about the lack of sensitivity of the IR spectral type on $\log~g$ is the strength of the FeH absorption feature at 0.99~$\mu$m.  In contrast to the synthetic spectra of the 4500~K stars, the spectra of the 3500~K stars both possess FeH, a result of the temperature sensitivity of this feature.  However, the two stars with different surface gravities have similarly strong FeH features both of which are clearly weaker than the FeH in the spectrum of DQ~Tau.  According to the study of \citet{schi1997}, FeH is strongest in the spectra of the coolest, dwarf stars (lowest temperature, highest surface gravity).  According to the comparison made in Figure~\ref{fig:tio_g}, the higher surface gravity of the $\log g$ = 4.5 is not enough to account for the strength of the FeH feature.  Figure~\ref{fig:tio_g} also demonstrates that a photospheric temperature of 3500~K is not {\it low enough} to produce an FeH feature as strong as is observed in DQ~Tau.  In order to simultaneously fit the TiO features and the FeH feature in DQ Tau, a synthetic spectrum with a lower photospheric temperature (cooler than an M2) and/or a slightly higher surface gravity (greater than an M2III) is required.  Therefore, we conclude that strengths of both the FeH and the TiO features indicate that a low surface gravity associated with the ``puffed up'' envelopes of contracting TTSs cannot account for the discrepancy between the infrared and optical spectral types for such sources.
 
 \subsubsection{Color Anomalies \& Star Spots}

\begin{figure*}[ht]
\begin{center}
\includegraphics[angle=0,width=1.8\columnwidth]{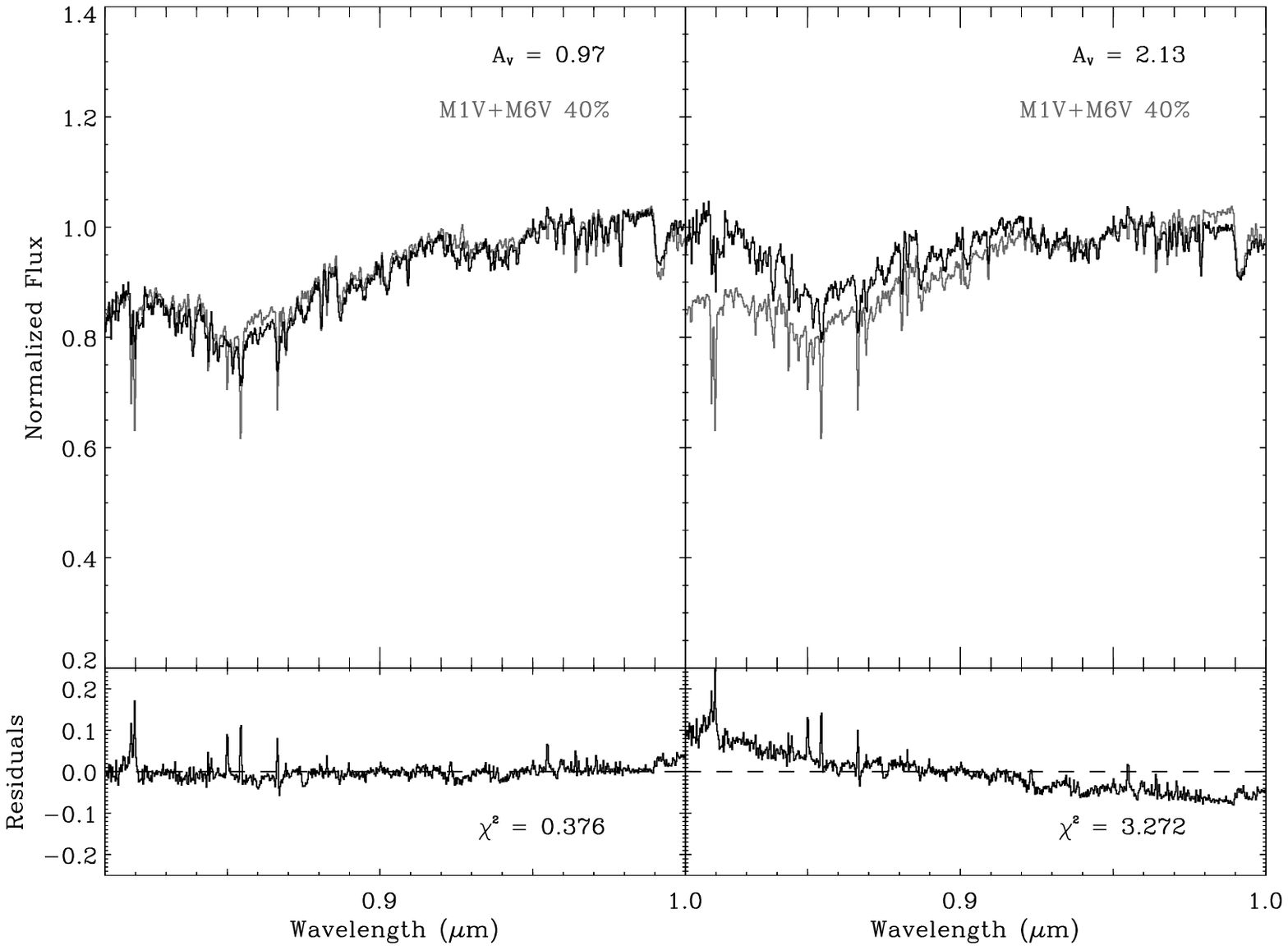}
\caption{Two panels showing fits of a spotted star to a DQ~Tau spectrum (2011 November 13 UT) dereddened with two different values of visual extinction: $A_V$~=~0.97 and 2.13.  The spotted star spectrum (gray) is a composite of an M1V standard representing the photosphere and a cooler M6V standard representing the spot with 40\% coverage.  Corresponding residuals are plotted in the adjacent, bottom panel.  The DQ~Tau spectrum with $A_V$~=~0.97 provides a superior fit to the spotted star spectrum.}
\label{fig:M6V_40}
\end{center}
\end{figure*}

\begin{figure*}[ht]
\begin{center}
\includegraphics[angle=0,width=1.8\columnwidth]{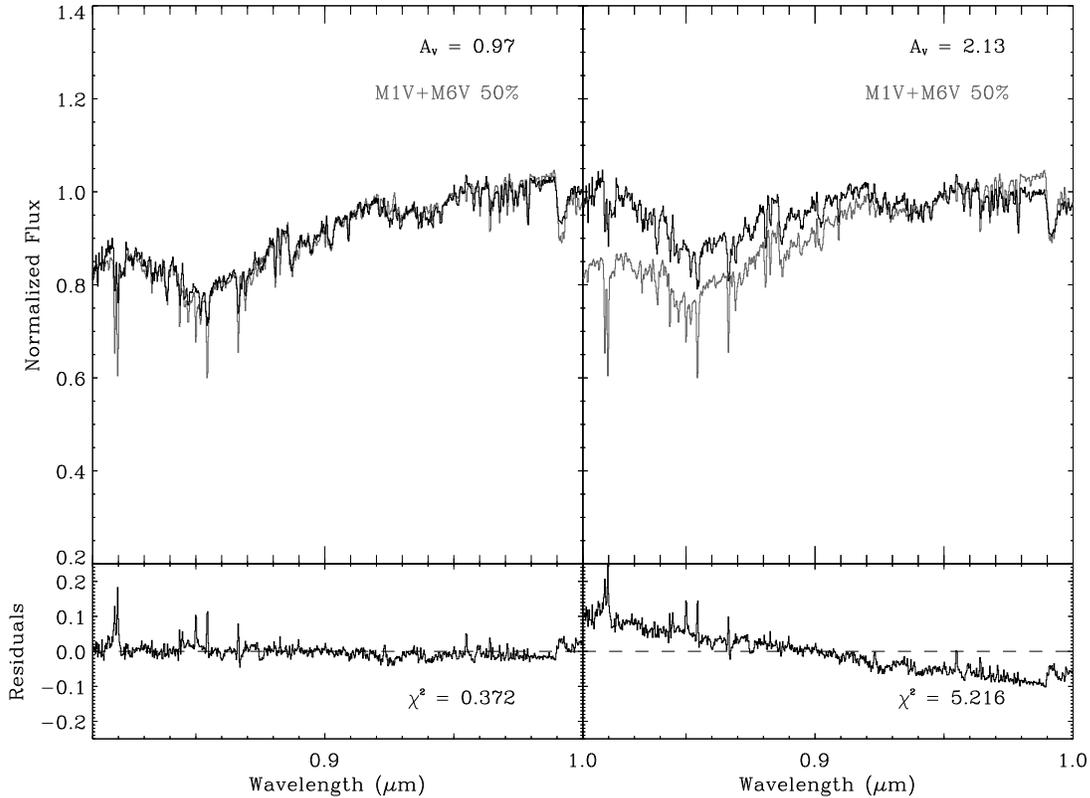}
\caption{Same as Figure~\ref{fig:M6V_40}, this figure shows a slightly better fit for a spot size of 50\%.}
\label{fig:M6V_50}
\end{center}
\end{figure*}

\begin{figure*}[ht]
\begin{center}
\includegraphics[angle=0,width=1.8\columnwidth]{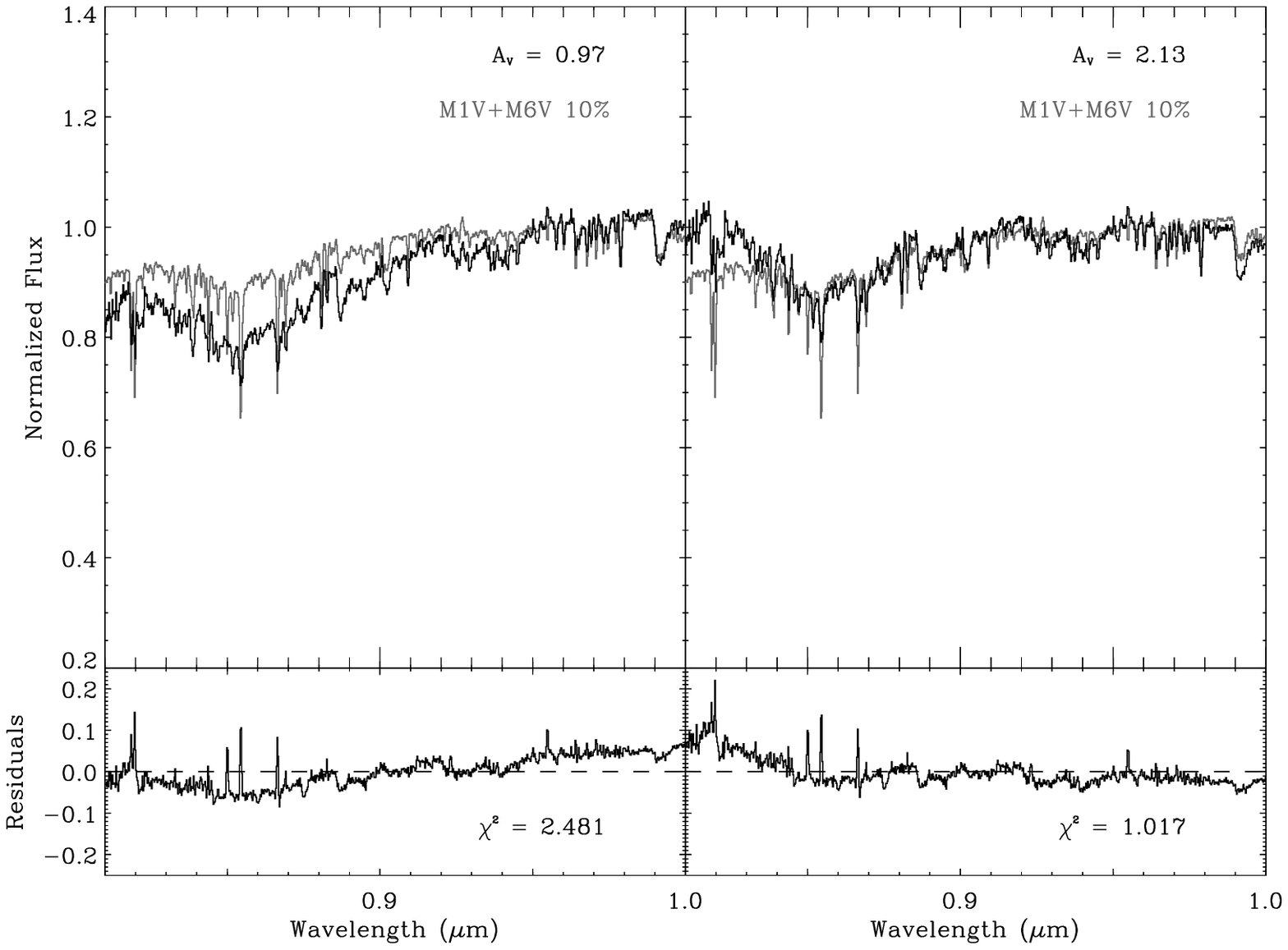}
\caption{Same as Figure~\ref{fig:M6V_40}, this figure shows a poor fit for a spot size of 10\% for $A_V$~=~0.97.  However, modestly good fits are found for the higher values of visual extinction.}
\label{fig:M6V_10}
\end{center}
\end{figure*}

Using infrared measurements, \citet{gull1998b} and \citet{fisc2011} demonstrated that weak-line or non-accreting TTSs appear brighter at longer wavelengths than main-sequence stars of the same spectral types.  One result of this color anomaly is that $A_V$ values determined from IR photometry will be larger than those determined from optical photometry.  \citet{gull1998,gull1998b} and \citet{fisc2011} find systematically larger values of $A_V$ for both non-accreting and accreting TTSs as compared to values determined from optical wavelengths.  Gullbring et al.\ also demonstrate that spectral types derived for their sample of wTTSs will be later types when determined from longer wavelengths, similar to what we find for DQ~Tau in the previous section \citep[see Figures~4 and 5 from;][]{gull1998b}.

We used near-IR photometry from \citet{keny1995} to calculate $A_V$ for DQ~Tau (M1V) and found $A_V$ = 3.6 and 6.6 from $E(J - K)$ and $E(H - K)$\footnote{We used $\frac{A_V}{{E(J - K)}}$~$=$~5.82 and $\frac{A_V}{E(H - K)}$~$=$~15.3, respectively.  Also note that the effect would have been magnified if we had assumed a spectral type earlier than M1V.}, respectively.  Dereddening the DQ~Tau spectra with such large values for $A_V$ would drastically affect the shape of the infrared continuum, forcing it to appear as a much earlier type star.  \citet{gull1998b} suggest that very large axisymmetric spots may be responsible for the color anomalies and differences in values of $A_V$. 
 
A defining characteristic of TTSs is their irregular photometric variability \citep{joy1945,joy1949}, with one of the well-established sources of variability being the rotation and evolution of large, cool star spots on their surfaces \citep{bouv1989,herb1994}.  The spot temperatures and fractional coverage of the stellar surfaces in TTSs are shown to be similar to those found on RS CVn stars \citep{hall1972,bouv1989}.  Considerable effort has been committed to understanding the sizes and temperatures of spots on the surfaces of magnetically active stars, perhaps, more so than have been carried out for TTSs.  Spot sizes, temperatures, and locations are deduced from photometric light-curve modeling and spectral line diagnostics \citep[i.e., Doppler imaging;][]{stra2009}.  Relying on the fact that spots on active, low surface gravity stars are similar to those on TTSs, we discuss findings from a few studies of the TiO molecular bands in the photospheres of TTSs and RS CVn stars that are relevant for the analysis we perform below.

\citet{neff1995}, \citet{onea1996}, and \citet{onea1998} used high-resolution spectroscopy of the TiO bands in the spectra of several RS CVn-type stars to constrain the spot and non-spot surface temperatures as well as the spot sizes.  In these three papers, the authors developed a technique for fitting the RS~CVn spectra using inactive G and K stars as spectral standards for the non-spotted photosphere and inactive M standards for the spots.  For late-type active stars, \citet{onea1998} found spot temperatures between 3400 and 3900~K and spot filling factors between 0.1 and 0.5 (10\% and 50\% fractional coverage, respectively), in good agreement with photometric light-curve modeling.  

\citet{herb1990} and \citet{berd1992} first showed that variations in the strengths of the TiO bands near 7100~\AA\ correlated with the existence, rotation, and evolution of cool star spots on the surfaces of TTSs.  In light of this correlation, the anomalous strengths of both the TiO and FeH absorption features in the spectra of DQ~Tau may strongly suggest the presence of cool spots on the surfaces of the DQ~Tau companions \citep{schi1997}.  Inspired by the evidence pointing towards the contribution of cool spots to the spectral type discrepancy, the inability of surface gravity effects to reproduce the strengths of the TiO and FeH features, and the spectral fitting technique developed by \citet{onea1998}, we attempted to fit the TiO features at 0.85 and 0.88~$\mu$m and the FeH feature at 0.99~$\mu$m to determine if cool spots could be responsible for the spectral type discrepancy observed for DQ~Tau.  Modeling of the entire near-IR spectra would require a disk model to account for the infrared excess emission beyond 1~$\mu$m.  Therefore, we focus on fitting the DQ~Tau spectra in the spectral window from 0.8 to 1.0~$\mu$m and reserve the inclusion of disk models and associated parameters to future study.

Main-sequence spectral standards from the SpeX library were used to construct composite spectra, representative of spotted stars.  We chose M0V and M1V standard spectra to represent the photosphere based on optically-derived spectral types.  M6V and M7V standards were chosen to represent the spot temperatures of 3050~K and 2650~K, respectively.  Composite spectra were calculated for six spot filling factors ($f_{sp}$$\sim$0.1, 0.2, 0.3, 0.4, 0.5, and 0.6).  The spot filling factors weighted the contributions of the spectra produced by the spots to the final composite spectra.  A goodness-of-fit to the observed DQ~Tau spectrum was determined by finding the composite spectrum with the minimum $\chi^2$-value for the two different photosphere and spot temperatures, six spot filling factors, and two of the published values of $A_V$: 0.97 and 2.13.

The best fits to the TiO features and the spectral shape between 0.8 and 1.0~$\mu$m was given by an M1V photosphere (M1V$_{ph}$) and a spot temperature of 3050~K (M6V$_{sp}$), with spot filling factors, $f_{sp}$~=~0.4 and~0.5, and $A_V$~=~0.97 (see Figures~\ref{fig:M6V_40} and \ref{fig:M6V_50}).  The $\chi^2$-values, which are calculated from the residuals for all of the data points in the spectral window, presented in Table~\ref{tbl:chisq} indicate that the M0V$_{ph}$+M6V$_{sp}$ model with $f_{sp}$~=~0.6 is the best fit.  While the fit of the spectral shape is slightly better, the fit to the FeH feature is extremely poor.  Therefore, we dismiss this model as the best overall fit to the DQ~Tau spectrum.  We also take the M1V$_{ph}$+M6V$_{sp}$ model with $f_{sp}$~=~0.4 to be a slightly better fit than the $f_s$~=~0.5 model based on the fitting of the FeH feature.  The FeH feature is best fit by the smaller spot size, $f_{sp}$~=~0.3 for the M1V$_{ph}$+M6V$_{sp}$ model.  Based on the work of \citet{schi1997}, we can interpret this discrepancy between fits to the TiO and the FeH as a combination of the strict low temperature sensitivity of the FeH feature combined with its dependence on $\log g$.  In other words, both a higher spot temperature and/or a smaller $\log g$ for the model star could account for the spot size discrepancy.  This is observed in the models with the hotter M0V photosphere as the larger spot size $f_{sp}$~=~0.4 produces the best fit to the FeH feature.

The atomic absorption features of Na~{\scshape ii} and Ca~{\scshape ii} infrared triplet are much weaker in the best-fit models as demonstrated by the residuals in Figures~\ref{fig:M6V_40} and \ref{fig:M6V_50}.   The Ca~{\scshape ii} absorption lines in the spectrum of DQ~Tau are likely filled in partially by Ca~{\scshape ii} emission related to the accretion activity in the system as well as continuum veiling.  The Na~{\scshape ii} features though less affected by accretion activity, will also be affected by any veiling at these shorter wavelengths.  We discuss difficulties associated with measuring veiling in this region of the spectrum in the following section.

In determining the significance of the fits, it is also important to note that for the larger value of $A_V$~=~2.13, the TiO bands and overall spectral shape in the 0.8-1.0~$\mu$m window were fit reasonably well by models with the smallest spot filling factor of 0.1 (see Figure~\ref{fig:M6V_10}).  However, the poor fit at the shortest wavelengths in this spectral region as well as the poor approximation of the FeH band by the smaller spot size result in the comparatively large $\chi^2$-values.  Even though the $\chi^2$-values steadily decrease with spot size for both M1V$_{ph}$ models, the poor approximation of the FeH feature leads us to conclude that spots with $f_{sp}$~$<$~0.1 are not viable models for DQ~Tau.  Table~\ref{tbl:chisq} presents the $\chi^2$-values for the twenty different combinations of spot sizes and photospheric and spot temperatures.

\begin{deluxetable*}{ccccc}
\tabletypesize{\scriptsize}
\tablecaption{$\chi^2$-values for Spotted Star Model Fits\label{tbl:chisq}}
\tablewidth{0pt}
\tablehead{
			  	\colhead{Spot Filling Factor}
			&  	\colhead{M0V$_{ph}$+M6V$_{sp}$}
			&  	\colhead{M1V$_{ph}$+M6V$_{sp}$}
			&      \colhead{M0V$_{ph}$+M7V$_{sp}$}
			&	\colhead{M1V$_{ph}$+M7V$_{sp}$} \\			
				\colhead{($f_s$)}
			&	\colhead{($A_V$=0.97, 2.13)}
			&	\colhead{($A_V$=0.97, 2.13)}
			&	\colhead{($A_V$=0.97, 2.13)}
			&	\colhead{($A_V$=0.97, 2.13)}}
\startdata
0.1  &  5.82, 1.17  &   2.48, 1.02  &  6.42, 1.32  &  2.86, 0.95 \\
0.2  &  4.32, 0.92  &   1.58, 1.37  &  5.42, 1.09  &  2.22, 1.10 \\
0.3  &  2.88, 0.94  &   0.84, 2.07  &  4.33, 0.93  &  1.58, 1.40 \\ 
0.4  &  1.59, 1.37  &   0.38, 3.27  &  3.16, 0.92  &  0.97, 1.95 \\
0.5  &  0.65, 2.46  &   0.37, 5.22  &  1.97, 1.21  &  0.51, 2.95 \\
0.6  &  0.35, 4.63  &   1.18, 8.27  &  0.92, 2.11  &  0.40, 4.75 \\
\enddata
\end{deluxetable*}

Based on the quality of the fits of the models possessing the larger spots and the smaller $A_V$ values of 0.97, we conclude that large, cool star spots make significant contributions to the near-infrared spectra of the DQ~Tau system and may partially account for the color anomaly and optical/infrared spectral type discrepancies previously observed for many TTSs.  

Such a conclusion is relevant for at least two recent studies that address the near-IR spectral types of TTSs.  First, \citet{vacc2011} present a detailed study of TW Hya, an accreting TTS in the TW Hydra Association \citep{kast1997}.  Based on the comparison of multiple atomic and molecular features and the strengths of the water absorption features at 1.4, 1.9, and 2.7~$\mu$m to spectral standards in the SpeX Library, Vacca \& Sandell revised the spectral type for TW~Hya from a K7V to an M2.5V.  The results of our study suggest that large star spots may account the spectral mismatch for this source as well, explaining why the molecular components and the near-IR spectral shape of TW~Hya would indicate a significantly later spectral type for the source.  A second study, \citet{mccl2013}, presented a detailed investigation of SpeX spectra of 10 accreting TTSs in the Taurus-Auriga star forming region.  \citet{mccl2013} used optical spectra to determine the spectral types and then went about fitting simultaneously for $A_V$, veiling (r$_\lambda$), and the IR continuum excess.  For stars with small spots, the results from this study should accurately describe both the stellar photospheres and IR continuum excesses associated with these sources.  However, should any of these stars have large star spots, we find that the determination of IR continuum excesses measured for these sources will be impacted, with the value of $A_V$ being most strongly affected.  Therefore, one must address the strength of the molecular features and the spectral shape of the IR continuum when solving for $A_V$, veiling, and continuum excess.

\subsection{Veiling \& Spectral Variability}

\begin{figure}[ht]
\begin{center}
\includegraphics[angle=90,width=1.0\columnwidth]{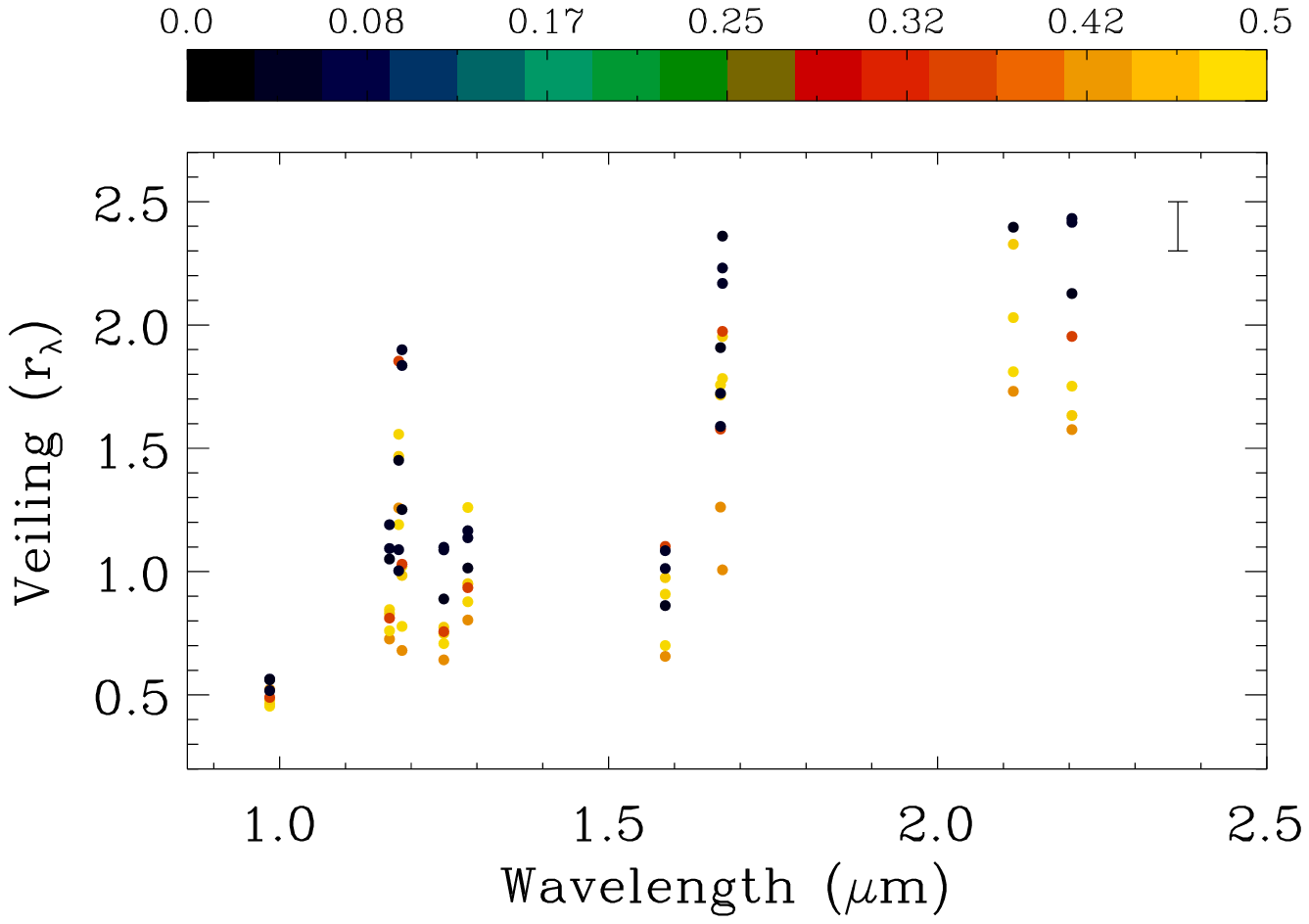}
\caption{Plot of veiling, r$_\lambda$, for eleven spectral features, including the molecular FeH feature for the eight epochs for which we have moderate resolution spectra (see data in Table~3).  The color coded data points identify veiling measurements with $\Delta\phi$, orbital phase difference from periastron (e.g., $\phi$~=~0.666 becomes $\Delta\phi$~=~0.334).  The general trend is that the veiling is highest for periastron passes ($\Delta\phi$~=~0.0), which correspond with the highest level accretion activity.  Note the small variation in the veiling measured by the FeH feature.}
\label{fig:LkCa3_veil}
\end{center}
\end{figure}

\begin{figure*}[ht]
\begin{center}
\includegraphics[angle=0,width=1.8\columnwidth]{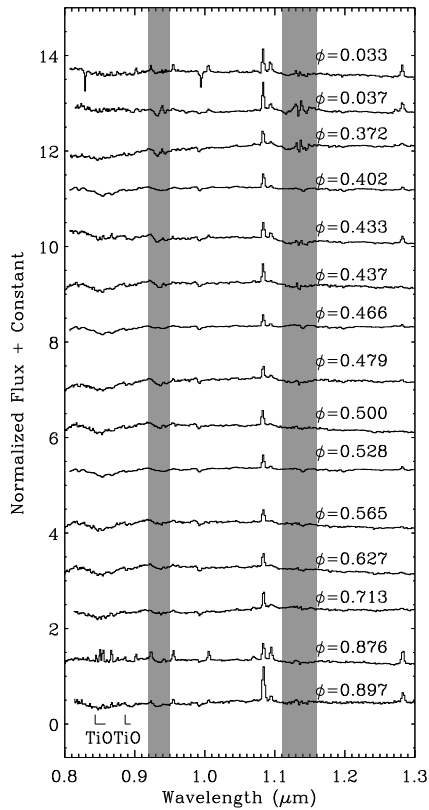}
\caption{Fifteen epochs of DQ~Tau spectra taken with CorMASS and SpeX, which include the TiO bands at 0.85 and 0.88~$\mu$m, are plotted from 0.8-1.3~$\mu$m.  Variations in both the strength of the TiO absorption bands and the shapes of the continua are clear.  We note that the TiO bands are weakest in epochs associated with accretion flares, indicated by the strongest H~{\scshape i} and Ca~{\scshape ii} triplet emission ($\phi$~=~0.033, 0.037, 0.372, 0.433, \& 0.876).}
\label{fig:tio_var}
\end{center}
\end{figure*}

\begin{figure*}[ht]
\begin{center}
\includegraphics[angle=0,width=1.8\columnwidth]{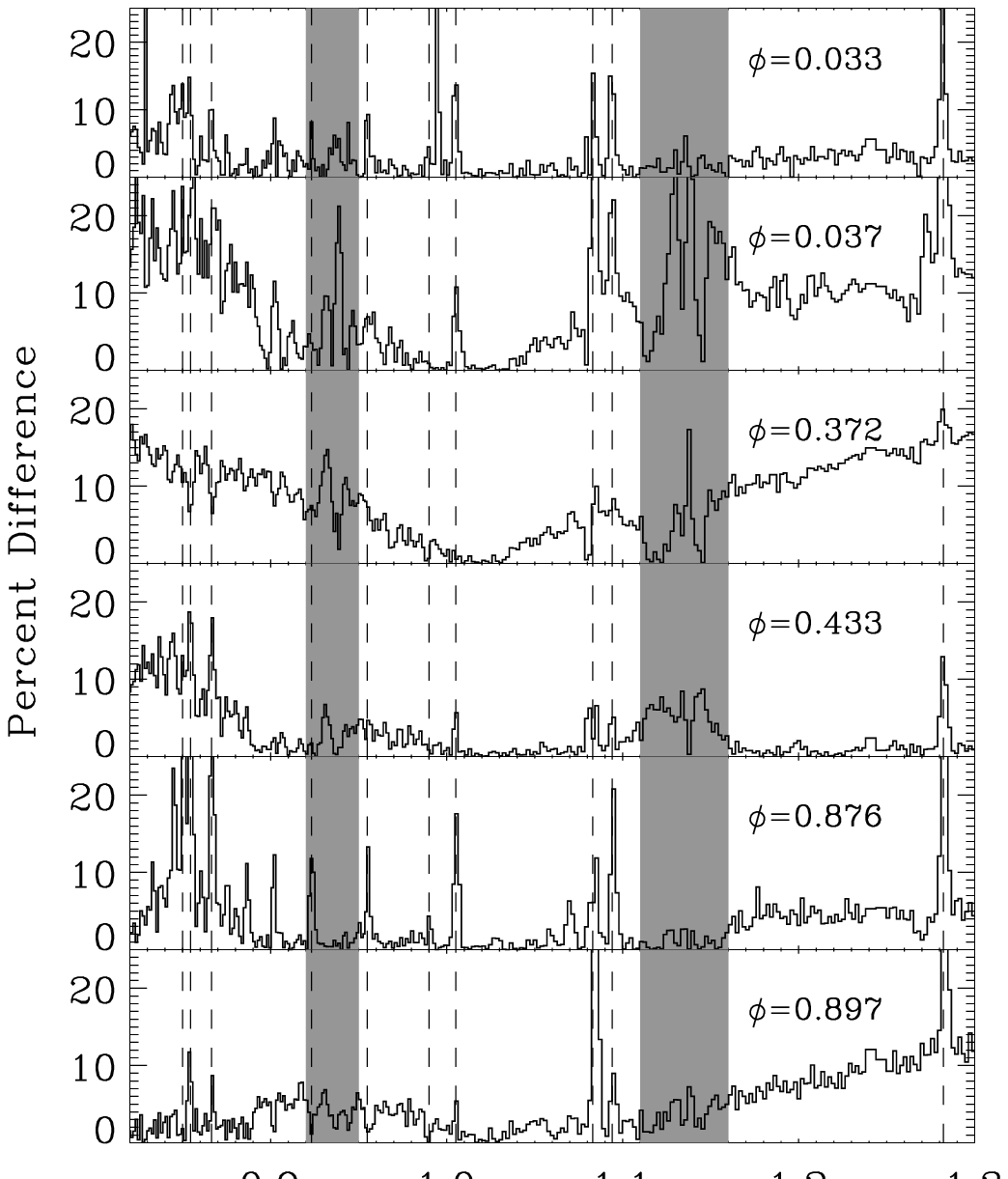}
\caption{Plots of the percent difference for five epochs of DQ~Tau over 0.8 to 1.3~$\mu$m in which the 2003 December 14 UT observation at $\phi$~=~0.500 was used as the reference spectrum.  Both the reference and target spectra were normalized to the continuum near 1.02~$\mu$m before the spectra were subtracted and then divided by the reference spectrum.  Variations in the shape of the continuum and strength of the predominant spectral features are clearly depicted.  The TiO feature experiences some of the most significant variability with the exception of the $\phi$~$=$~0.897 near periastron.  We note that the TiO bands are weakest in epochs associated with accretion flares, indicated by the strongest H~{\scshape i} and Ca~{\scshape ii} triplet emission ($\phi$~=~0.033, 0.037, 0.372, 0.433, \& 0.876).}
\label{fig:var}
\end{center}
\end{figure*}

Having established both a spectral type and visual extinction, we proceed to measure the veiling as a function of wavelength.  We follow a standard measurement procedure based on equivalent widths of several temperature sensitive metallic lines in the dereddened spectrum and derive the veiling without taking spots into account \citep{fisc2011,vacc2011,mccl2013}.  Note that we do not simultaneously solve for veiling, visual extinction, and infrared excess as \citet{fisc2011} and \citet{mccl2013}.  Instead, we assume an M1V spectral type, $A_V$~=~0.97, and no infrared excess. The veiling measurements for all epochs with moderate spectral resolution are presented in Table~\ref{tbl:3} and Figure~\ref{fig:LkCa3_veil}.   Based on the strengths of the accretion indicators, we know that the magnitudes of the periastron accretion flares vary from passage to passage.  Therefore, it is not surprising to see in Figure~\ref{fig:LkCa3_veil} that the veiling is not always the highest at periastron for all lines measured.  The same can be said for the veiling outside periastron passage where the values also fluctuate and indicate that, for some lines, the veiling can remain high through much of the orbit.  Despite the range of values measured for veiling at periastron, a general trend suggests that veiling is consistently higher at periastron than the rest of the orbit.  

\begin{deluxetable*}{lccccccccccc}
\tabletypesize{\footnotesize}
\tablecaption{Veiling Measurements by Orbital Phase\label{tbl:3}}
\tablewidth{0pt}
\setlength{\tabcolsep}{0.0in}
\tablehead{
				\colhead{Orbital}
			&  	\colhead{r$_{\rm FeH}$}
			&	\colhead{r$_{\rm K I}$}
			&	\colhead{r$_{\rm Mg I}$}
			&	\colhead{r$_{\rm Fe I}$}
			&	\colhead{r$_{\rm Mg I}$}
			&	\colhead{r$_{\rm Fe I}$}
			&	\colhead{r$_{\rm Mg I}$}
			&	\colhead{r$_{\rm Al I}$}
			&	\colhead{r$_{\rm Al I}$}
			&	\colhead{r$_{\rm Al I}$}
			&	\colhead{r$_{\rm Na I}$}  \\
				\colhead{Phase}
			&	\colhead{0.99~$\mu$m}
			&	\colhead{1.169~$\mu$m}
			&	\colhead{1.183~$\mu$m}
			&	\colhead{1.189~$\mu$m}
			& 	\colhead{1.244~$\mu$m}
			&	\colhead{1.264~$\mu$m}
			&	\colhead{1.588~$\mu$m}
			&	\colhead{1.673~$\mu$m}
			&	\colhead{1.676~$\mu$m}
			&	\colhead{2.117~$\mu$m}
			&	\colhead{2.206~$\mu$m}}
\startdata
0.402		&   0.52    &  0.73  &  1.25  &  0.68  &  0.64  &  0.80  &  0.66  &  1.26  &  1.01  &  1.73  &  1.58   \\
0.466		&   0.45    &  0.76  &  1.19  &  0.78  &  0.71  &  1.26  &  0.70  &\nodata&\nodata& 1.81 &   0.04   \\
0.539		&   0.46    &  0.85  &  1.56   & 1.02  &  0.75  &  0.88  &  0.91  &  1.76   & 1.78    & 2.03  &  1.75  \\
0.546		&   0.48    &  0.83  &  1.47   &  0.98  & 0.77  &  0.95  &  0.98  &  1.72   &  1.95  &  2.33   &  1.63  \\
0.666		&   0.49    &  0.81  &  1.85   &  1.03  &  0.76 &  0.93  &  1.10  &  1.58   &  1.97  &\nodata&  1.95  \\
0.961		&   0.56    &  1.09  &  1.09   &  1.84  &  1.10  &  1.17 &  1.01  &  1.59   &  2.23  &\nodata&  2.43  \\
0.962		&   0.52    &  1.19  &  1.00   &  1.90  &  1.09  &  1.14 &  1.08  &  1.72   &  2.17  &\nodata&  2.42  \\
0.963		&   0.56    &  1.05  &  1.45   &  1.25  &  0.89  &  1.01 &  0.86  &  1.91  &  2.36   &  2.40   &  2.13  \\
\enddata
\end{deluxetable*}

The formation of the molecular absorption features within cool spots indicates the potential for variability in the strengths of the 0.85 and 0.88~$\mu$m TiO and 0.99~$\mu$m FeH bands in DQ~Tau due to the rotation of the stars and evolution of the spots \citep{herb1990,berd1992}, in addition to veiling of the features and the potential for heating during accretion events.  In Figure~\ref{fig:tio_var}, we present 15 epochs of DQ~Tau spectra for which we have data on both the TiO and FeH bands, plotting the percent difference between a DQ~Tau reference spectrum taken at apastron ($\phi$~=~0.500) during an apparent quiescent accretion phase and each of the spectra that show clear evidence for variability in both the continuum shape and strength of spectral features.  Visual inspection of these spectra reveal significant variations in the strength of the TiO features, but less so for the FeH feature.  The shape and depth of the features are roughly consistent for observations outside of periastron passes, with the exception of $\phi$~=~0.372 and 0.433.  In general, the largest variations in the strengths of both the atomic and molecular features coincide with epochs when the star is experiencing an increase in accretion activity.  The TiO bands weaken significantly during epochs in which the accretion indicators are in or near a flare state, including the two exceptional flare epochs taken near apastron.  However, one epoch very near periastron ($\phi$~$=$~0.897) shows little evidence of variation from the reference spectrum and thus retains a strong TiO band.  Interestingly, the increase in strength of the Ca~{\scshape ii} and H~{\scshape i} features should correlate with a ``filling'' in of the TiO features due to veiling associated with the accretion activity.

The majority of the data that includes both the TiO and FeH bands are low resolution CorMASS spectra, making difficult a simultaneous measurement of the veiling in both the TiO and FeH features.  Establishing a continuum level in order to measure the strength of the TiO features also increases the uncertainty in such a measurement.  In addition, a measurement of the strength of the TiO band is complicated by the contribution of the Ca~{\scshape ii} emission features residing within the TiO band, which increases the uncertainty in the measurement of the TiO feature and the estimate of its variability.  Assuming that the variation in the strength of the feature is due solely to veiling and after smoothing over the calcium features, we estimate r$_{\rm TiO}$ to be $\sim$0.4 during epochs in the quiescent phase of the orbit ($\phi$~=~0.404, 0.460, and 0.530) for which we have SpeX observations,  and conservatively estimate a lower limit of r$_{\rm TiO}$~$\sim$0.6 and an upper limit of $\sim$1.0 for the heightened accretion activity occurring in the periastron passages observed with CorMASS\footnote {Note that TSpec does not offer wavelength coverage of the TiO features}.  This is a significantly larger range of veiling than those measured for FeH.  If both features, as suggested by the spot models, originate in large cool spots, we would expect them to vary in concert with one another.  However, we do expect the FeH to be most strongly associated with the spots given its substantially lower dissociation energy of 1.6~eV as compared to the 6.8~eV for the TiO molecule.  Unfortunately, the limited spectral resolution of our current data set limits our ability to draw a meaningful conclusion about the variability of these features.  Future data sets with greater resolving power and high signal to noise will be important for probing the sources of variability such as veiling, spot evolution, spot heating, and stellar rotation.

\subsection{H$_2$ Emission from Accretion Shocks}

\citet{hart1995} and \citet{basr1997} observe forbidden line emission with features possessing both blue- and red-shifted velocity components characteristic of a disk wind.  However, we find no evidence in the literature for a molecular outflow associated with DQ~Tau binary.  In a recent high-spatial resolution spectral imaging study of GG~Tau~Aa and Ab, another binary system that possesses standard indicators of outflowing gas \citep{hart2003}, but no evidence for a molecular outflow, \citet{beck2012} find evidence for shocked molecular hydrogen emission ($v$~$=$~1-0~S(1) at 2.1218~$\mu$m) residing inside the inner truncation radius of the circumbinary disk in this system.  While this study lacks the spectral resolution necessary to determine if the gas is confined to an accretion streamer or an outflowing wind, previous high resolution long-slit spectroscopy of the 2.1218~$\mu$m feature constrains the bulk motion of the gas to within 1~km~s$^{-1}$ of systemic velocity for GG~Tau~A \citep{bary2003}.  With a full width at half maximum of $\leq$14~km~s$^{-1}$, it is not likely that the molecular gas is entrained in a fast moving bipolar outflow.  Therefore, Beck et al.\ conclude that the H$_2$ gas most likely traces shocked molecular gas resulting from accretion streamers falling inward from the circumbinary disk.

If this novel stimulation mechanism is responsible for the emitting H$_2$ gas in GG~Tau~A, it may also be capable of producing similarly excited H$_2$ emission in a system like DQ~Tau.  Uniquely, in the case of DQ~Tau, the dynamics of the system provides predictable accretion flares that may result in variable emission from shocked H$_2$ confined to the accretion streams.  Given the strength of the H$_2$ emission detected in the GG~Tau~A system, we searched the only moderate resolution spectrum taken of DQ~Tau at periastron on 2 October 2012 UT for the 2.12~$\mu$m H$_2$ emission.  Not surprisingly, without a high resolution spectrum, the feature was not detected and an upper limit of 6.0~$\times$~10$^{-14}$~ergs~s$^{-1}$~cm$^{-2}$ was measured.  Indeed, at this sensitivity, the GG~Tau~A H$_2$ emission feature ($F_{2.12}$~$=$~6.9~$\times$~10$^{-15}$~ergs~s$^{-1}$~cm$^{-2}$) would not have been detected.  We plan to observe DQ~Tau at high-spectral resolution during a periastron passage in the future to further test the idea of shocked gas as part of accretion streamers in binary systems.

\section{Summary}

Using a rich spectroscopic dataset, we have explored many aspects of the DQ~Tau binary system.  We leveraged the fact that the variations in accretion activity are orbitally-modulated and thus predictable to study accretion in this system, detecting what appears to be a surprising accretion flare phased with an apastron passage in the system.  Such an enhancement of the accretion activity likely suggests that the inner radius of the circumbinary disk is not smooth.  Predictions of irregularly-shaped, non-axisymmetric structures extending inwards from the circumbinary disk provide a reasonable explanation for the nature and rarity of the apastron flares.  Apastron flare events could be an important step towards constraining the viscosity of the circumbinary disk.  A dedicated monitoring program will be needed to characterize such accretion events.  The applicability of orbitally-modulated accretion activity to young single-star systems possessing a forming giant planet in a circumstellar disk highlights the importance for continuing the study of accretion in this context.  With respect to measuring the physical conditions of the accreting gas, we find that higher spectral resolution data with high signal-to-noise ratios are needed.  Dedicated monitoring of DQ~Tau and similar tight, accreting binary systems is exactly what is needed to build up on the insights gained from this study.

We also have demonstrated that large star spots have the potential to account for several aspects of the optical/IR spectral type mismatches previously observed for TTSs, but often overlooked.  Given the impact on the shape of the IR spectral and the strength of several molecular features, large cool star spots should be included as parameters when modeling the infrared spectra of TTSs.  With the acquisition of broader wavelength coverage spectra with higher resolution, we plan to build upon the rudimentary models presented here by expanding the parameter space for photospheric and spot temperatures and adding veiling effects and basic disk models.

\acknowledgments

The authors would like to thank Tracy Beck, Suzan Edwards, Bill Herbst, Praveen Kundurthy, Sean Matt, James Muzerolle, Matt Nelson, Dawn Peterson, Ricardo Schiavon, Mike Skrutskie, Bill Vacca, and John Wilson for contributions and useful discussions regarding many aspects of this work.  We thank David Whelan for collecting one of the TripleSpec observations.  MSP acknowledges partial funding support from the Division of Natural Sciences at Colgate University through the summer research program.  Thanks to David Weintraub for a close reading and comments on the initial manuscript.  We also thank the telescope operators from Apache Point Observatory, Russet McMillan and Jack Dembicky, as well as Bill Golisch at the IRTF.  Also thanks to Sean Matt and Sacha Brun for graciously hosting JSB's visit to CEA Saclay.  JSB recognizes the seminal contributions to the study of spotted stars made by his former teacher and mentor, Douglas S. Hall.  The authors thank the anonymous referee for useful comments and suggestions regarding this work.  This study made use of SAO/ADS and the Simbad database.

 \end{document}